\documentclass[review]{elsarticle}

\usepackage{hyperref}

\usepackage{amssymb}
\usepackage{bm}
\usepackage{color}
\usepackage{mathrsfs}
\usepackage{natbib}
\usepackage{enumerate}
\usepackage{geometry}
\usepackage{graphics}

\journal{Journal of \LaTeX\ Templates}

\biboptions{authoryear}

\def\ga{\ \lower 3pt\hbox{${\buildrel > \over \sim}$}\ }
\def\la{\ \lower 3pt\hbox{${\buildrel < \over \sim}$}\ }
\def\mm{M_{\rm m}}
\def\mp{M_{\rm p}}
\def\rs{R_{\rm m}}
\def\rp{R_{\rm p}}
\def\qm{Q_{\rm m}}
\def\qp{Q_{\rm p}}
\def\km{k_{\rm 2m}}
\def\kp{k_{\rm 2p}}

\begin{document}

\begin{frontmatter}

\title{Orbital evolution of Saturn's mid-sized moons and the tidal heating of Enceladus}

\author[myadress]{Ayano Nakajima\corref{cor1}}
\ead{nakajima.a.ah@m.titech.ac.jp}

 \author[secadress]{Shigeru Ida}
 
 \author[thirdadress]{Jun Kimura}
 
 \author[secadress]{Ramon Brasser}
 
\cortext[cor1]{Corresponding author}
 

\address[myadress]{Department of Earth and Planetary Sciences, Tokyo Institute of Technology, 2-12-1 Ookayama, Meguro-ku, Tokyo 152-8551, Japan}
\address[secadress]{Earth-Life Science Institute, Tokyo Institute of Technology, 2-12-1 Ookayama, Meguro-ku, Tokyo 152-8550, Japan}
\address[thirdadress]{Department of Earth and Space Science, Osaka University, 1-1 Machikaneyama-cho, Toyonaka City, Osaka 560-0043, Japan}

\begin{abstract}
The formation and orbital evolution of Saturn's inner mid-sized moons -- Rhea, Dione, Tethys, Enceladus, and Mimas -- are still 
debated. The most puzzling aspects are 1) how the Tethys-Dione pair and the Mimas-Enceladus pair passed through 
their strong 3:2 mean-motion resonances during the tidal orbital evolution, and 2) the current strong heat flow from Enceladus, which 
is a few orders of magnitude higher than the tidal energy dissipation caused by the present orbital eccentricity of Enceladus. 
Here we perform N-body simulations 
of the moons' orbital evolution from various initial conditions -- assuming that the moons were formed 
from Saturn's hypothetical massive ring --
and investigate possible paths to solve the above difficulties. 
If the moons remain on nearly circular orbits and the influence of the rings is neglected, 
we find that the Tethys-Dione pair cannot avoid becoming trapped in the 2:1 and 3:2 mean-motion 
resonances as they recede from Saturn,
and that the Tethys-Enceladus pair cannot avoid collisions after the resonance trapping, 
in case Saturn's quality factor is smaller than 15\,000.
These findings are inconsistent with the current orbital configuration. 
However, taking into account both the eccentricity excitation and the orbital expansion caused by the ring torque, we find that these resonance captures are avoided. 
With the relatively high eccentricity pumped up by the torque,
Enceladus passes through all the mean-motion resonances with Tethys,
and the Dione-Tethys pair passes through their 2:1 resonance
and possibly the 3:2 resonance as well.
After Enceladus resides beyond the 2:1 resonance with the outer ring edge,
the eccentricity can be tidally damped.
While this is a promising path of evolution, in most runs,
Enceladus collides with Tethys by the excited eccentricity.
There is a hint that a ring mass decrease 
(possibly due to Mimas formation)
could avoid the collision between Enceladus and Tethys.
The parameter survey taking into account detailed ring evolution
and Mimas is left for future study.
The heat that was tidally dissipated due to the eccentricity excitation by the
ring torque in the past is stored in the moons and slowly radiated away through conductive transfer. 
The stored heat in Enceladus may account for the current anomalously high heat flow.

\end{abstract}

\begin{keyword}
 Saturn \sep moons \sep resonance \sep Enceladus
\MSC[2010] 00-01\sep  99-00
\end{keyword}

\end{frontmatter}


\section{Introduction and tidal evolution}
\label{sec: 1}
The evolution and origin of Saturn's mid-sized moons -- Mimas, Enceladus, Tethys, Dione, and Rhea -- remain an enigma. 
Located closer than Saturn's massive moon Titan, but farther away than Saturn's famous ring system and a collection of much 
smaller moons, the classical mid-sized moons form a rich dynamical system both now and in the past.

The masses and orbital elements of these moons are listed in Table \ref{tab:orb}. 
When ignoring mutual gravitational interactions and orbital eccentricities, the relative tidal expansion rate of each 
moon's semi-major axis ($a$) is given by \citep[e.g.,][]{MurrayDermott1999,Goldreich1966}

\begin{eqnarray}
\frac{1}{a}\frac{d a}{dt}= 3 \frac{\kp}{Q_{\rm p}}
\frac{\mm}{\mp} \left(\frac{R_{\rm p}}{a}\right)^5 \Omega,
\label{eq:a_evol}
\end{eqnarray}
where $\Omega = \sqrt{G(\mp+\mm)/a^3}$ is the orbital frequency of each moon, and $G$, $M$ and $R$ are the gravitational constant, mass and physical radius, respectively. $Q_{\rm p}$ and $\kp$ are the quality factor and Love number of the host planet (Saturn). Hereinafter, the subscript ``m'' and ``p'' indicate moon and planet. 

\begin{figure}[htbp]
  \centering
 \includegraphics[width=120mm, angle = 0]{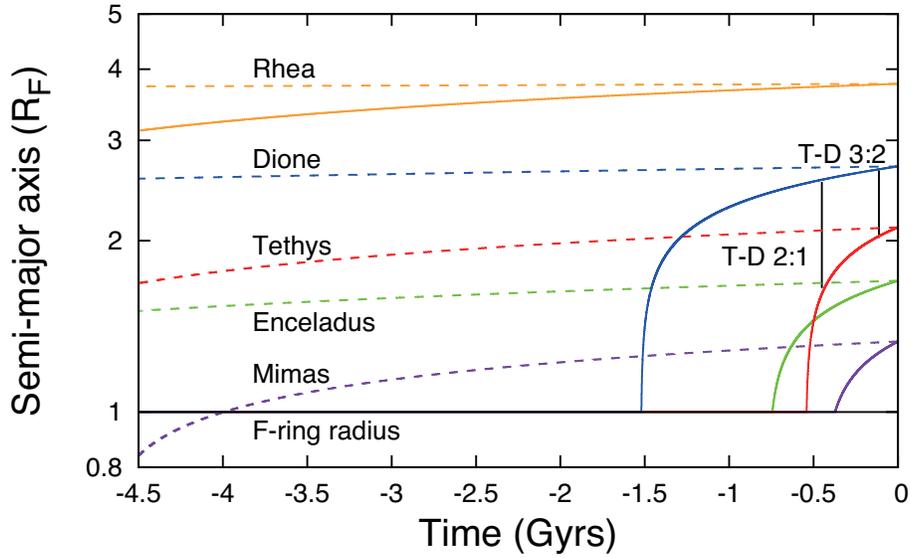}
  \caption{Evolution of semi-major axis of Mimas, Enceladus, Tethys, Dione, and Rhea,
  backwardly integrating Eq.~(\ref{eq:a_evol}) from the current semi-major axes without mutual interactions between the moons. 
  The dashed lines assumed $Q_{\rm{p}} = 18\,000$ and the solid lines assumed $Q_{\rm{p}} = 1\,700$ which are normalized by the F-ring radius ($R_{\rm F} = a_{\rm{F}} = 1$). We assumed Saturn's Love number $\kp = 0.34$ for all the moons in both lines. The vertical black lines indicate the location of most recent 3:2 and 2:1 mean motion resonances between Tethys and Dione.}
  \label{fig: kaiseki}
\end{figure}

Saturn's $k_{\rm 2p}$ is often assumed to be in the range of 0.3-0.4 \citep{Gavrilov1977,Helled2013}.
On the other hand, the quality factor of Saturn is not so well determined.
Previously, it was often thought that $Q_{\rm p} \sim 18\,000$ is a lower bound inferred by assuming that Mimas 
migrated from the outside of the synchronous radius to its current orbit over the past 4.5 Gyr \citep{Peale1980a}. 
Recently, however, \citet{Lainey2012,Lainey2017} proposed a much lower, and controversial, value $Q_{\rm{p}} = 1\,700 
\pm 500$, which was deduced from the detailed analysis of astrometric observational data of the tidal evolution of Saturn's major 
moons. This lower value of $Q_{\rm p}$ implies a much faster tidal evolution. Figure \ref{fig: kaiseki} 
shows a backward integration of Eq.~(\ref{eq:a_evol}) for each moon starting from their current orbits down to the F-ring.
Hereafter, we scale the semi-major axis of the moons by the average distance of the F-ring to Saturn ($a_{\rm F} \simeq 1.40 
\times 10^5, {\rm km} \simeq 2.4 \rp$ where $\rp$ is Saturn's physical radius), which is comparable to the planet's Roche limit 
($r_{\rm Roche} = 2.4\rp \left(\rho_{\rm p}/\rho_{\rm m}\right)^{1/3}$, where $\rho_{\rm p}$ and $\rho_{\rm m}$ are the bulk density of Saturn and material, respectively).

We used both $\qp = 18\,000$ and $\qp = 1\,700$, 
assuming $\kp = 0.34$ \citep{Gavrilov1977} for all the moons.
We use the same value of $\kp/\qp$ for all the mid-sized moons in our simulations. 
For Rhea, \citet{Lainey2017} suggested that $k_{\rm 2p}/Q_{\rm p}$ is ten times smaller than we assumed.
In our simulation, we mostly focus on Enceladus, Tethys and Dione. 
Our assumption that all the moons have the same $\kp/Q_{\rm p}$ does not change the result.

The lower value of $Q_{\rm p}$ has profound implications for the formation of Saturn's inner moons, 
not least of which is that they can no longer be primordial. 
One theory for the formation of the moons was proposed by \citet{Charnoz2011}, 
who suggested that these mid-sized moons were formed relatively recently from the spreading of a previous massive ring rather than 
from an extended circumplanetary protosatellite disk \citep[also see][]{Crida2012,Salmon2017}. 
The theory suggested by \citet{Crida2012} makes predictions about 
the mass and semi-major axis relationship of the moons that accurately fit their observed distribution, 
and $Q_{\rm p}$ needs to be small enough for Saturn's mid-sized moons to be formed from the spreading 
of massive rings within the age of the Solar System.
Recently \citet{Fuller2016} proposed that $Q_{\rm p}$ gradually decreased from 
the large values in an early phase of resonant locking between Saturn's oscillation mode and the moon's orbital frequency
in the course of Saturn's interior evolution and tidal orbital evolution. In this case, the mid-sized moons were formed from an 
extended circumplanetary protosatellite disk over 4.5 Gyr ago and the tidal orbital expansion was slow in the
early phase until resonant locking occurred. 
Further research is needed to distinguish between the two cases, though they aren't mutually exclusive.

\begin{table}
\begin{center}
\begin{tabular}{c|ccc} \hline
                 & semi-major axis ($a_{\rm F}$) & mass ($10^{-6} M_{\rm p}$) & eccentricity \\ \hline \hline 
Rhea         & 3.77  & 4.07 & 0.0010 \\
Dione        & 2.70  & 1.94 & 0.0022 \\
Tethys       & 2.11 & 1.09 & 0.0000 \\
Enceladus & 1.70  & 0.190 & 0.0045 \\
Mimas       & 1.33  & 0.067 & 0.0202 \\ \hline
\end{tabular}
\end{center}
\caption{Physical parameters of Saturn's mid-sized moons.
The semi-major axis is scaled by the orbital radius of the F ring ($a_{\rm F} \simeq 1.40 \times 10^5 \, {\rm km}$)
and the masses are scaled by $10^{-6}$ times Saturn's mass $\mp$. 
These values are cited from NASA Space Science Data Coordinated Archive.}
\label{tab:orb}
\end{table}

In the case that $(1/a)(da/dt)$ of an outer moon is smaller than that of an inner moon,
the migration is termed ``convergent", while it is ``divergent" otherwise.
Two or more moons can be captured in a mean-motion resonance if
the migration is convergent and ``adiabatic", i.e. when the characteristic libration timescale of the resonance angle is much 
shorter than the migration timescale across the resonant width \citep[e.g.,][]{MurrayDermott1999}.
Currently, Tethys is just inside of the 3:2 resonance with Dione and their migration is convergent. Figure 1 shows that the 
Tethys-Dione pair should have passed a strong 3:2 resonance and also their 2:1 resonance if $\qp \la 15\,000$.
If their eccentricities remained low throughout their migration, 
it is not yet clear how they avoided or escaped from these resonances. 
Even if the time-dependent $Q_{\rm p}$ modeled by \citet{Fuller2016} is considered, 
the convergent migration near the 3:2 resonance does 
not change and the problem of capture in the 3:2 resonance cannot be easily avoided.

Even when considering the formation of these moons, the resonance capture problem remains. 
\citet{Salmon2017} performed N-body simulations of collisional growth of the mid-sized moons, 
based on \citet{Charnoz2011}'s model. 
While they have succeeded to reproduce the overall mass-distance distribution of the mid-sized moons, 
the capture of Tethys and Dione in the 3:2 resonance still looks to be a setback in their results. 
\citet{Zhang2012} pointed out the possibility that 
a large ($\approx 250$km diameter) and slow ($\approx 0.5$km/s) impact that created the Odysseus basin on Tethys
could have knocked the Tethys-Dione pair out of their 3:2 $e$-Dione resonance.
If we consider such an impact which may occur during the satellite formation, 
the resonance can be broken and reproduce the current orbit,
although the occurrence probability of such an impact is not clear. 

For relatively small $\qp$, Fig.~\ref{fig: kaiseki} suggests that Enceladus was formed earlier than Tethys and Tethys overtook Enceladus 
during their tidal migration. Note that Fig.~\ref{fig: kaiseki} does not include the effect of mutual interactions between the moons.
In section 3.1, we will show that such an orbital crossing cannot occur because of the resonant interaction between Enceladus and 
Tethys. We shall further point out that this orbital crossing is not required if we include the additional orbital expansion 
by ring torque, which is very rapid near the ring's outer edge. 
On the other hand, its force vanishes beyond the 2:1 resonance with the edge of the ring
\citep[e.g.,][]{Charnoz2011,Crida2012}. We will also point out that eccentricity excitation by the ring torque 
\citep[e.g.,][]{Goldreich2003b,Duffell2015} may need to be taken into account in addition to its effect on the orbital expansion, although 
the eccentricity excitation was not considered in previous simulations \citep{Charnoz2011,Crida2012,Salmon2017}.
If the eccentricity is excited over a threshold value, the capture probability at a resonance is significantly reduced 
\citep[e.g.,][]{Malhotra1993} and a pair of moons can avoid becoming trapped.

The high surface heat flux of Enceladus, which is observed to be $\sim 15.9 \pm 3.1$~GW 
in the South Polar Terrain by the Cassini CIRS instrument \citep{Howett2011}, 
is also a big mystery. 
If Enceladus is in a thermal equilibrium, the heat flux must be equal to the heat production in its interior.
However, it is unlikely that Enceladus is in thermal equilibrium. The present radiogenic heating within the rocky core is estimated 
to be only $\sim 0.3$ GW \citep{Porco2006}. The tidal energy dissipation rate is \citep{MurrayDermott1999}
\begin{eqnarray}
H = \frac{21}{2} \Omega \frac{G \mp^2}{a} \frac{\km}{\qm} \left(\frac{\rs}{a}\right)^5 e^2 \sim 0.1 
\left(\frac{\km/\qm}{10^{-4}}\right) \left(\frac{\rs}{R_{\rm E}}\right)^5 \left(\frac{a}{a_{\rm E}}\right)^{15/2} \left(\frac{e}{e_{\rm 
E}}\right)^2 \,{\rm GW},
 \label{eq:HEnceladus}
\end{eqnarray}
where $a_{\rm E}, e_{\rm E}$ (see Table 1) and $R_{\rm E}$ ($\simeq 250\,{\rm km}$) 
are semi-major axis, eccentricity ($e$) and physical radius ($R$) of Enceladus (the subscript ``E" represents Enceladus) and we used Saturn's mass for $M_{\rm p}$. 
\footnote{
This formula applies for a moon in a synchronous rotation.
If libration is taken into account, tidal dissipation becomes stronger 
\citep[][also see the comment below]{Ferraz-Mello2017}.
}
Even with relatively high tidal damping parameters in Enceladus of $\km/\qm \sim 10^{-4}$, 
the estimated tidal heating is only $\sim 0.1$ GW at the current 
small eccentricity of Enceladus ($\simeq 0.0045$). 
\citet{Meyer2007} estimated the total tidal energy dissipation for Enceladus and 
Dione at the current 2:1 resonance as $H \simeq 1.1 \left(18\,000/Q_{\rm p}\right)$ GW for $\km/\qm \sim 10^{-4}$, 
assuming that the system is in a steady state. 
If the moons are in a steady state, 
$e^2$ in Eq.~(\ref{eq:HEnceladus}) is proportional to $k_{\rm{2p}}/Q_{\rm p}$ (see Appendix A).
Since tidal heat production is inversely proportional to $Q_{\rm{p}}$, the heating is as high as $\sim 
10$ GW for $Q_{\rm{p}} \sim 2\,000$. However, such great heating requires 10 times larger $e$ than the current value of Enceladus, as 
shown in Eq.~(\ref{eq:HEnceladus}), unless the moon damping parameters are extremely high ($\km/\qm \sim 10^{-2}$).
Although the extremely dissipative case of Enceladus
is not completely ruled out \citep[e.g.][]{Ferraz-Mello2017,Choblet2017},
a very strong damping would prevent the moons from passing 
mean-motion resonances as discussed in
section \ref{subsec:both}.

This inconsistency strongly suggests that the current heat flux from the surface is not equilibrated with the current heat production 
in Enceladus.
\citet{Ojakangas1986} proposed the idea that Enceladus' eccentricity increases most of the time with
episodical decreases, generating a large amount of heat; the current state
is the end of the high heat generation phase with a fully damped eccentricity. 
However, \citet{Meyer2008b} argued that such oscillation does not occur
with the Ojakangas-Stevenson model.
\citet{ONeill2010} proposed an episodic heat release. On occasion, the generated heat is stored in the interior, and subsequently the 
stored heat is episodically released. 
This idea suggests that we are observing the narrow window of a
high heat energy release from Enceladus. 

Here we consider the possibility that the intense heat production during the past orbital evolution of Enceladus is the result 
of eccentricity excitation due to the ring torque, and that this heat is still stored in Enceladus' interior and is slowly
being released.
Enceladus' potential past high eccentricity caused by the suggested ring torque is damped after it migrates beyond the 2:1 
resonance with the ring edge, as we will clearly show in Section \ref{subsec:both}. Since the tidal heat rate is proportional to $e^2$
and the ring torque easily excites $e$ to values of $\sim$0.05-0.1, the stored heat can be large enough to account for the current high 
heat flux. With the recently proposed $Q_{\rm p} \sim 2\,000$, the high $e$ orbital phase of Enceladus could be recent enough for Enceladus 
to keep the stored heat in its interior. 
Note that a close encounter between two moons can only
pump the eccentricity of the smaller body up to a value that depends on
the surface escape velocity ($v_{\rm esc}$)  of the larger one.
The corresponding eccentricity is
$\sim v_{\rm esc}/v_{\rm K} = [2(\mm/\mp)(a_{\rm m}/\rs)]^{1/2}
\sim 2 [(\mm/\mp)^{2/3}(a_{\rm m}/a_{\rm F})]^{1/2} \sim 0.02$ where $v_K$ is Keplerian velocity
and $a_{\rm F}$ is comparable to the Roche radius,
$a_{\rm F} \sim r_{\rm Roche} \sim 2.4\left(\rho_{\rm p}/\rho_{\rm m} \right)^{1/3} \rp
\sim 2.4\left(\rho_{\rm p}\rp^3/\rho R_{\rm m}^3 \right)^{1/3} R_{\rm m} \sim
2.4\left(M_{\rm p}/M_{\rm m} \right)^{1/3} R_{\rm m}$.
The past high-eccentricity phase is not the result of past close encounters, but was instead caused 
by resonant interactions between moons or pumping by the ring torque.

Here, we investigate the orbital evolution of the mid-sized moons, based on the model of formation from the spreading ring. We 
employ N-body simulations to compute the evolution and discuss resonance capture and tidal heat production
due to eccentricity evolution. We take into account the changes in eccentricities and semi-major axes of
the moons according to tides in Saturn and the moons themselves. We also perform runs where we include orbit torques caused by the 
ring, including potential eccentricity excitation.
The formation models of \citet{Charnoz2011} and \citet{Crida2012} were semi-analytical and did not calculate the gravitational 
interactions between the moons, including resonant configurations and eccentricity evolution.
The N-body simulations by \citet{Salmon2017} were the first to investigate the formation scenario of \citet{Charnoz2011} and 
\citet{Crida2012} in more detail. 
Although they reproduced the overall mass distribution of these moons, they did not discuss the details of resonance capture/break-up or eccentricity evolution in individual systems, both of which are important to discuss the current orbital architecture of the moons.

We focus our investigation on resonance passing/capture/break-up and tidal heat production with N-body simulations,
and therefore we do not include early collisional growth of moonlets from the spreading ring.

The outline of our paper is as follows. 
Section \ref{sec: methods} describes our numerical model, and Section \ref{sec: results}  presents our results of the orbital 
evolution of the moons and discusses the resonance trapping in detail. 
In Section \ref{sec: heat estimation} we estimate the heat energy of Enceladus stored in the course of orbital evolution. 
Finally, we present our conclusions and a discussion in the last section.

\section{Methods}
\label{sec: methods}
\subsection{Numerical model}

We simulate the orbital evolution of the system -- which mainly consists of Saturn, Enceladus, Tethys, and Dione -- to investigate the detailed orbital evolution of strongly
interacting moons starting from many different initial conditions;
in some runs we also added Rhea.
Dynamically, Rhea is almost decoupled from other moons.
Mimas has the smallest mass and could not affect other moons' motions significantly.

Figure~\ref{fig: kaiseki} suggests that the Enceladus-Tethys pair undergoes
orbital crossing if $Q_{\rm p} \la 15\,000$.
According to the recently proposed smaller $\qp$ \citep{Lainey2012,Lainey2017}, 
we adopt constant $\qp \sim 2\,000$-4\,000 in our simulations.
Because the backward integration does not include
gravitational perturbations --such as the resonant perturbations-- between the moons, 
the backward integration cannot be available in the entire space from the birth places of the moons
to their current positions.
Therefore, we need to perform forward N-body simulations
with various initial conditions
to investigate which initial conditions result in the current orbital configuration
by making individual simulations simple.
We will explain the initial conditions in more detail in section \ref{subsec:param}.

We have incorporated the semi-major axis expansion and the eccentricity damping due to tidal interaction 
into the N-body code SyMBA \citep{Duncan1998}. 
In our simulation, we accelerate tidal evolution by increasing the Love numbers of both the planet and the moons by the same factor, $C = 10^3 - 10^4$, to reduce computation time. 
This ``speed-up factor" has been used in many other works (e.g. \citealp{Malhotra1990,showman1997,Meyer2008a,Zhang2009}).
The ratio between tidal orbital expansion and eccentricity damping rates
is kept the same for different values of the speed-up factor, in order to maintain consistency
(see discussions below).
Furthermore, the time in all of following figures represents real time that is obtained by simulation time multiplied by the speed-up factor ($C$).
The orbital expansion and eccentricity excitation by the ring torque are implemented in the N-body simulations,
and the acceleration with the same speed-up factor for tides is applied.
Some aspects in orbital changes by mutual gravitational interactions,
including resonant interactions, cannot be accelerated
because they are calculated by N-body simulation.
We will discuss the effects of the speed-up factor on the probability of
resonant trapping.

In our simulations, we find that collisions between moons
occur after orbital eccentricities are excited by resonant perturbations
or the ring torque.
Consequently, the collision velocity ($v_{\rm col}$) is usually larger than the surface escape velocities ($v_{\rm esc}$) of the moons
and the collisions are usually ``hit-and-run" collisions \citep{Asphaug2006}.
We use the model by \citet{Genda2012} based on SPH simulation results.
The critical collision velocity ($v_{\rm{cr}}$), so that a collision between body 1 and 2 (their masses are 
$M_1$ and $M_2$) is hit-and-run, 
is given by
\begin{equation}
\frac{v_{\rm{cr}}}{{v_{\rm{esc}}}} = c_1\Gamma^2\Theta^{c_5} + c_2\Gamma^2 + c_3\Theta^{c_5} + c_4,
\end{equation}
where $\Gamma = | M_1 - M_2 | / (M_1 + M_2)$, $\Theta = 1 - \sin{\theta}$ ($\theta$ is impact angle), and $v_{\rm{esc}}$ is escape velocity. 
The fitting parameters are $c_1 = 2.43$, $c_2 = -0.0408$, $c_3 = 1.86$, $c_4 = 1.08$ and $c_5 = 2.50$ \citep{Genda2012}. 
If $v_{\rm{col}} < v_{\rm{cr}}$, we assume that the collision results in merging,
while it is hit-and-run otherwise.
In this study, we consider the most optimal case to reproduce the current orbits,
preserving the moons against catastrophic disruption.
For simplicity, in the case of hit-and-run collisions, 
we have the moons pass through each other in a softened gravitational potential
without collisional energy dissipation.
We find that once moons start orbit crossing, they undergo repeated
hit-and-run collisions, where we assume that they eventually coalesce,
regardless of whether or not we include the energy dissipation during the hit-and-run collision.

\subsection{Tidal forces}%

Tidal deformation of the host planet caused by the moons
transfers angular momentum from the planetary spin to the moon's orbit.
To express the orbital expansion rate in Eq.~(\ref{eq:a_evol}), 
the tangential force per unit mass 
is added to the equations of motion, which is given by
\begin{eqnarray}
f_{\rm p,\psi} & \simeq & \frac{1}{a} \frac{d\sqrt{ GM_{\rm p} a }}{dt} 
= \frac{1}{2a} \frac{da}{dt} \sqrt{ \frac{GM_{\rm p}}{a} } \nonumber \\
 & = & \frac{3}{2} \frac{k_{\rm 2p}}{Q_{\rm p}}\frac{G\mm}{a^2} \left(\frac{R_{\rm p}}{a}\right)^5 
 \simeq \frac{3}{2}\frac{k_{\rm 2p}}{Q_{\rm p}} \frac{G\mm}{r^2}\left( \frac{R_{\rm p}}{r} \right)^5,
 \label{eq:fpsi}
\end{eqnarray}
where $r$ means the orbital radius of the moons.
In our simulation, we set the radial component of the planetary tidal force to be zero, 
because it does not affect the orbital expansion.

A moon's deformation caused by the planet dissipates the moon's orbital kinetic energy,
which results in a decrease in the moon's eccentricity ($e$) and semi-major axis ($a$).
The eccentricity damping timescale is given by \citep[e.g.,][]{MurrayDermott1999}
\begin{eqnarray}
\frac{1}{\tau_e} = -\frac{1}{e}\frac{de}{dt}  =  
\frac{21 \km}{2 \qm}
\frac{M_{\rm p}}{\mm} \left(\frac{\rs}{a}\right)^5 \Omega.
\label{eq:e_damp}
\end{eqnarray}
For the current orbital elements of Enceladus and
$k_{2m}/Q_m \sim 10^{-5}$, $\tau_e \sim 8 \times 10^8$yrs.
Using the eccentricity damping timescale, we add
the following tidal force per unit mass
caused by the moon's deformation
in a simple form to the equations of motion \citep{Kominami2002},
\begin{eqnarray}
\bm{f}_{\rm m} = - \frac{\bm{v}-\bm{v}_{\rm{K}}}{\tau_{\rm{e}}},
\label{eq:fm}
\end{eqnarray}
where $\bm{v}$ is the moon's velocity and $\bm{v}_{\rm{K}}$ is the local circular Keplerian velocity. 
Adding the semi-major axis damping associated with the eccentricity damping, 
the semi-major axis changes as
\begin{eqnarray}
\frac{1}{a}\frac{d a}{dt}= 3 \frac{\kp}{\qp}
\frac{\mm}{M_{\rm p}} \left(\frac{R_{\rm p}}{a}\right)^5 \Omega
- \frac{21 \km}{\qm}
\frac{M_{\rm p}}{\mm} \left(\frac{R_{\rm m}}{a}\right)^5 e^2 \Omega.
\label{eq:a_evol2}
\end{eqnarray}
The second term in Eq. (\ref{eq:a_evol2}) corresponds to $-2e^2/\tau_e$ in Eq.~(\ref{eq:a_evol2}).
Because the moon's orbital angular momentum is conserved during the eccentricity damping,
$0 = (1/L) dL/t = (1/2a) da/dt - (e/(e^2-1)) de/dt 
\sim a/(2 \tau_a) - e^2/\tau_e.$ 
We do not include the second term in $f_{\rm p,\psi}$
in Eq.~(\ref{eq:fpsi}),
because it is automatically caused by $\bm{f}_{\rm m}$ (Eq.~(\ref{eq:fm})).
The second term is dominant when
\begin{equation}
e \ga \left[ 
\frac{1}{7} \frac{\kp/\qp}{\km/\qm} 
\left(\frac{M_{\rm m}}{M_{\rm p}}\right)^2
\left(\frac{R_{\rm p}}{R_{\rm m}}\right)^5
\right]^{1/2}
\sim 0.1 \left(\frac{\kp/\qp}{10^{-4}}\right)^{1/2}
\left(\frac{\km/\qm}{10^{-5}}\right)^{-1/2}
\left(\frac{M_{\rm m}/M_{\rm p}}{10^{-6}} \right)^{1/6}.
\label{eq:a_evol3}
\end{equation}

\subsection{Saturn's ring torque}%

In one set of runs (SET2), we include the torque from Saturn's ring following \citet*{Crida2012}.
The timescale is given by
\begin{eqnarray}
\frac{1}{\tau_{a, \rm ring}}
 = \frac{1}{a}\frac{d a}{dt} 
 = \frac{a_{\rm F}}{a}  \frac{d (\Delta a/a_{\rm F})}{dt} 
 =\frac{16}{27 \pi}\frac{M_{\rm ring}}{M_{\rm p}} \frac{\mm}{M_{\rm p}}
\left( \frac{\Delta a}{a_{\rm F}} \right)^{-3} \frac{a_{\rm F}}{a} \, \Omega,
\label{eq:ring_a}
\end{eqnarray}
where $\Delta a = a - a_{\rm F}$ is the separation from F-ring, $M_{\rm ring}$ is Saturn's ring mass, which may be comparable to the moon's mass \citep{Crida2012} 
(see discussion below).
Note that this formula is an approximate one for computational simplicity and
the actual torque decreases in a discrete manner \citep[e.g.,][]{Meyer-Vernet1987}.
Because there is no 1st-order Lindblad resonance
beyond 2:1 resonance with the outer edge of the ring,
we need to introduce this discreteness at least beyond the 2:1 resonance.
We set the ring torque to vanish at $a \ga 1.59a_{\rm F}$.

\citet*{Goldreich2003b} and \citet*{Duffell2015} argued that 
the non co-orbital Lindblad torque excites eccentricity with the timescale as follows, 
\begin{equation}
\frac{1}{\tau_{e, \rm ring}} = \frac{1}{e}\frac{de}{dt} \,
\simeq \frac{1}{2 \pi}\frac{M_{\rm ring}}{M_{\rm p}}\frac{\mm}{M_{\rm{p}}} 
\left( \frac{\Delta a}{a_{\rm F}} \right)^{-4} \Omega,
\label{eq:ring_e}
\end{equation}
where we used $M_{\rm ring} \simeq \pi \Sigma a^2$.
In our simulation, we include the ring torque effects
on semi-major axis and eccentricity evolution in similar ways to
Eqs.~(\ref{eq:fpsi}) and (\ref{eq:fm}).

While the inner/outer Lindblad torques excite $e$,
the co-orbital Lindblad torque and corotation resonance torque damp $e$.
For a uniform surface density distribution without a gap,
the damping is stronger.
In the case of a gas disk, gas surface density is not zero even at the gap center.
Whether $e$ is excited or not depends on how deep the gap is.
On the other hand, ring particles are almost completely empty
outside the ring, so that $e$ of satellites should be excited.
\citet*{Goldreich2003b} and \citet*{Duffell2015} considered
a gas giant planet that opens up a gap in a protoplanetary disk.
Both inner and outer Lindblad torques excite $e$.
Because only Lindblad torque from an inner disk exists for the ring torque,
we decrease the formula by \citet*{Goldreich2003b} by a factor of 2. 

The $e$-excitation by the ring torque was not taken into
account in previous studies of Saturn's system \citep{Crida2012,Salmon2017}.
As will be shown later, the newly incorporated $e$-excitation
plays a key role in bypassing strong mean-motion resonances.

Note that Eqs.~(\ref{eq:ring_a}) and (\ref{eq:ring_e})
cannot be valid for $\Delta a \rightarrow 0$,
because ring particles are strongly scattered by the moon
for $\Delta a \la 2\sqrt{3}\, r_{\rm H}$, where $r_{\rm H}$ is the Hill radius of the moon 
defined by $r_{\rm H} = (\mm/3M_{\rm p})^{1/3} a \simeq 0.007 (\mm/10^{-6}M_{\rm p})^{1/3} a$
\citep{Ida1989}.
We should either start integration from $\Delta a > 2\sqrt{3}\, r_{\rm H}$
or start integration from $\Delta a = 0$ with adding a softening parameter 
$\epsilon \sim 2\sqrt{3}\, r_{\rm H}$ to $\Delta a$ in an approach 
 in which collisions are approximated with a softened gravitational potential.
Although $\epsilon \sim 2\sqrt{3}\, r_{\rm H}$ is physically justified,
a softening parameter is also often introduced to secure 
numerical stability and it corresponds to the numerical resolution.
By this reason, we here adopt $\epsilon = 10\, r_{\rm H}$.
Because this size is still much smaller than the typical migration distance of
satellites, this choice does not change the results.  

Equation~(\ref{eq:ring_e}) suggests that the initial eccentricity of
a satellite is important for the subsequent eccentricity evolution.
When satellites are formed from the ring, smaller clumps would interact with one another before coagulation to the satellites and the formed satellites interact with density fluctuations in the ring edge. The interactions are chaotic and an orbital eccentricity of ${\rm a \,few \,} \times r_{\rm H}/a$ would be excited when the orbital separation between the clumps or that between the satellite and the ring edge is smaller than ${\rm a \,few \,} \times r_{\rm H}$ \citep[e.g.,][]{Petit1986,Ida1990}, although N-body simulation would be required to prove this argument. 
Since $r_{\rm H}/a \simeq 0.007 (\mm/10^{-6}M_{\rm p})^{1/3}$,
we set the initial values of $e$ as $e_{0} \sim 0.01-0.03$.

We assume a constant ring mass ($M_{\rm ring}$) throughout each run for simplicity.
The assumed ring masses in individual runs are shown in Table 2.
The assumption of the constant $M_{\rm ring}$ does not affect our results, 
because 1) the decrease in $M_{\rm ring}$ would not be 
significant during the evolution we simulated and
2) the $a$-$e$ evolution path driven by the ring torque is 
independent of $M_{\rm ring}$.
The ring mass would largely change when new satellites are formed,
while it also changes gradually through viscous diffusion.
Here, we do not create a new satellite during the individual runs.
In SET2 runs, we start our simulations from the timing of the birth of Enceladus;
Dione and Tethys are already formed.
The ring mass should decrease at the time of the formation of Mimas, 
which we do not include in our simulations. 
The $a$-$e$ evolution path driven by the ring torque is independent of 
the speed-up factor $C$ and $M_{\rm ring}$, 
because $C$ is multiplied to
both Eq.~(\ref{eq:ring_a}) and Eq.~(\ref{eq:ring_e}) in the simulations
and both equations are proportional to $M_{\rm ring}$.
As we will show, the key process to reproduce the current orbital configurations
is to avoid capture by mean-motion resonances.
The capture can be avoided by sufficiently large $e$ and/or 
fast migration (non-adiabatic migration) of satellites. 
The $e$ evolution path as a function of $a$ 
is independent of $M_{\rm ring}$ and $C$ in the phase dominated by the ring torque.
The migration speed is proportional to $M_{\rm ring}$.
However, we will show that satellites are trapped at a resonance
that they encounter at the first time even with $C=10^3-10^4$,
which means that the capture is not changed even by order of magnitude larger 
values of $M_{\rm ring}$ in real cases with $C=1$.
Thus, the assumption of the constant ring mass and the artificial acceleration with $C$ 
are not critical 
flaws for the purpose of this paper, 
while simultaneous evolution of orbits and
the ring is important and left for future work.

The phase dominated by the ring torque is determined as follows.
Comparing Eq.~(\ref{eq:ring_a}) with the first term of Eq.~(\ref{eq:a_evol2}),
we find that the ring torque is dominant for 
\begin{eqnarray}
\frac{\Delta a}{a_{\rm F}}  <\frac{(\Delta a)_{{\rm crit,}a}}{a_{\rm F}} & \equiv & 
\left[ \frac{16}{81 \pi} \frac{M_{\rm{ring}}}{M_{\rm p}} \frac{Q_{\rm p}}{\kp} \left(\frac{a}{R_{\rm p}}\right)^4 \right]^{1/3} \\
  & \simeq & 
0.37 \left(\frac{\kp/\qp}{10^{-4}}\right)^{-1/3} \left(\frac{M_{\rm ring}/M_{\rm p}}{10^{-6}}\right)^{1/3} \left(\frac{a}{a_{\rm F}}\right)^{4/3},
 \label{eq:ring_dominate}
 \end{eqnarray}
where $a_{\rm F}$ is the F-ring radius ($a_{\rm F} \simeq 2.4 R_{\rm p}$).  
If we use $\kp/\qp \sim 10^{-4}$ and $M_{\rm ring}/M_{\rm p} \sim 10^{-6}$, 
the solution to Eq.~(\ref{eq:ring_dominate}) with $a = a_{\rm F} + (\Delta a)_{{\rm crit,}a}$
is $(\Delta a)_{{\rm crit,}a}\simeq 0.82 \, a_{\rm F}$. 
Because $\Delta a \simeq 0.59 \, a_{\rm F}$ at the 2:1 resonance with the ring edge, the ring torque is dominant in the orbital expansion until it diminishes at the 2:1 resonance as founded by \citet[][section 6.3]{Crida2012}.
Since $C$ is multiplied to
both Eq.~(\ref{eq:a_evol2}) and Eq.~(\ref{eq:ring_a}) in the simulations,
$(\Delta a)_{{\rm crit,}a}$ is independent of the value of $C$.
It depends on $M_{\rm ring}$, but only weakly. 
The eccentricity excitation competes with 
the tidal eccentricity damping.
It dominates over the tidal eccentricity damping,
as long as $\tau_{e, \rm ring} < \tau_e$, that is, 
\begin{eqnarray}
\frac{\Delta a}{a_{\rm F}}  <\frac{(\Delta a)_{{\rm crit,}e}}{a_{\rm F}} 
& \equiv &
\left[ \frac{2 \times 16}{21 \times 27 \pi} 
\frac{M_{\rm{ring}}}{M_{\rm p}} \left(\frac{M_{\rm m}}{M_{\rm p}}\right)^2 \frac{Q_{\rm m}}{k_{\rm 2m}} \left(\frac{a}{R_{\rm m}}\right)^5
\right]^{1/4} \\
& \simeq & 
0.2 \left(\frac{k_{\rm 2m}/Q_{\rm m}}{10^{-5}}\right)^{-1/4}
 \left(\frac{M_{\rm ring}/M_{\rm p}}{10^{-6}}\right)^{1/4} 
 \left(\frac{M_{\rm m}/M_{\rm p}}{10^{-6}}\right)^{1/12} \left(\frac{a}{a_{\rm F}}\right)^{5/4}.
 \label{eq:ring_e_dominate}
\end{eqnarray}
For $\km/\qm \sim 10^{-5}$ and $M_{\rm ring}/M_{\rm p} \sim M_{\rm m}/M_{\rm p} \sim10^{-6}$, 
the solution to Eq.~(\ref{eq:ring_e_dominate}) with $a = a_{\rm F} + (\Delta a)_{{\rm crit,}e}$
is $(\Delta a)_{{\rm crit,}e}\simeq 0.27 \, a_{\rm F}$. 
Again, $(\Delta a)_{{\rm crit,}e}$ is independent of the value of $C$
and it depends on $M_{\rm ring}$ only weakly.

\subsection{Initial conditions}
\label{subsec:param}
We performed four sets of simulations (SET1A, SET1B, SET2A, SET2B). 
Figure~\ref{fig: kaiseki} suggests that Enceladus formed earlier than
Tethys and they underwent orbit crossing, if we neglect the ring torque.
Because backward integration is not available
back to the state before the orbit crossing,
we examine many different initial conditions with
$a_{\rm{T,0}} < a_{\rm{E,0}}$ where the subscripts  ``T" and ``E" represent Tethys and Enceladus, 
and ``0" represents the initial values of the simulation.
We call this set of runs ``SET1A". 
We also carry out ``SET1B" where
Enceladus and Tethys are formed almost simultaneously 
in a horseshoe orbit, $a_{\rm{T,0}} \sim a_{\rm{E,0}}$.
As we will show in the next section,
many of these simulations produce a collision between Enceladus and Tethys, 
and therefore cannot reproduce
the current orbital configurations of Enceladus and Tethys.

\begin{figure}[htbp]
  \centering
  \includegraphics[width=70mm, angle = -90]{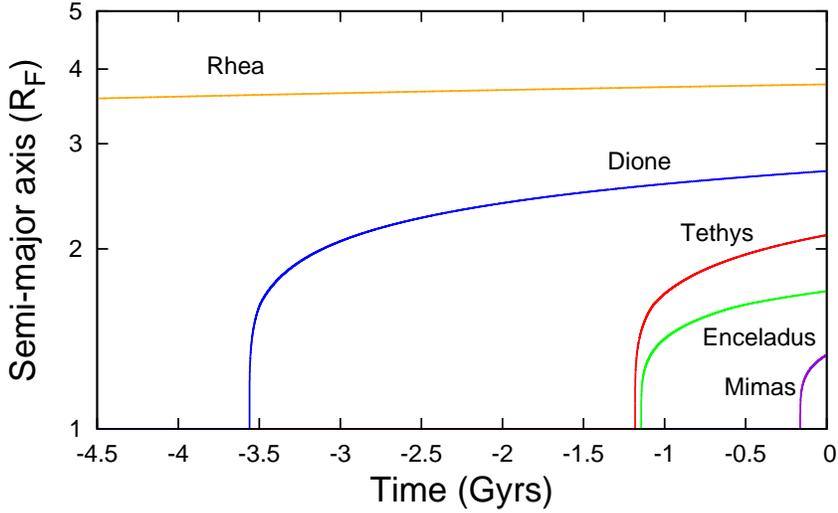}
  \caption{ The same as Figure~\ref{fig: kaiseki} except
  taking into account the torque from the ring.
  $\qp = 4000$ is used.
  The orbits of Mimas, Enceladus, Tethys, Dione, and Rhea
  are represented by violet, green, red, blue and orange lines.
  The initial mass of the ring is the sum of the masses of Dione, Tethys and 4$\times$Enceladus, respectively,
  and when one moon is swept out from the ring, the ring loses some of its own mass. }
  \label{fig:tide_evol_ring}
\end{figure}

As we already pointed out, the torque from the ring
would affect orbital evolution.
Figure \ref{fig:tide_evol_ring} shows backward tidal orbital evolution 
with $\qp = 4\,000$, taking into account the torque from the ring.
Because the evolution includes formation of Dione, Tethys, Enceladus and Mimas,
we changed the ring mass for this particular plot, while we use a constant ring mass
in N-body simulations.
The initial mass of the ring is the sum of the masses of Dione, 
Tethys and 4 $\times$ Enceladus, and when one moon is swept out from the ring, the ring loses some of its own mass.
Gravitational interactions between the moons are also neglected in this backward integration. 
For these parameters, 
it is suggested that Enceladus formed later than Tethys
and they can avoid the orbit crossing.
Because the orbital evolution by the ring torque is very rapid near the ring, 
it is likely that 
the rapid orbital evolution of Tethys had already ended when Enceladus began to form.
Accordingly, the Enceladus-Tethys pair's migration is convergent
in its early phase until Enceladus
migrates beyond the 2:1 resonance with the ring edge.
Tethys-Dione migration is always convergent.
Because the resonant and secular perturbations among these moons 
--which are not taken into account in Figure \ref{fig:tide_evol_ring}-- are complicated,
we also need to test various initial conditions
with $a_{\rm{T,0}} > a_{\rm{E,0}}$.
We call this set of runs ``SET2."
We will perform runs with only semi-major axis expansion by the ring torque (SET2A)
and others with both semi-major axis and eccentricity increases by the ring torque (SET2B).
We will show that only the runs with both semi-major axis and eccentricity increasing by the ring torque in SET2B potentially reproduce
the current orbital configurations of the mid-sized moons.

Initial masses of the moons are the same as the current masses.
In the simulations here, $\kp$, $\km$, $Q_{\rm p}$ and $\qm$ are 
set to be constant with time for all the moons and 
we adopt $\kp=0.34$, $\km=10^{-3}$, and $\qm=100$ in all runs.
We use $\qp = 1\,700 - 4\,000$ and the speed-up parameter of $C = 10^3-10^4$,
depending on runs.

\section{Results}
\label{sec: results}

\subsection{SET1: Enceladus forms no later than Tethys}
\begin{figure}[htbp]
  \centering
  \includegraphics[width=100mm, angle = 0]{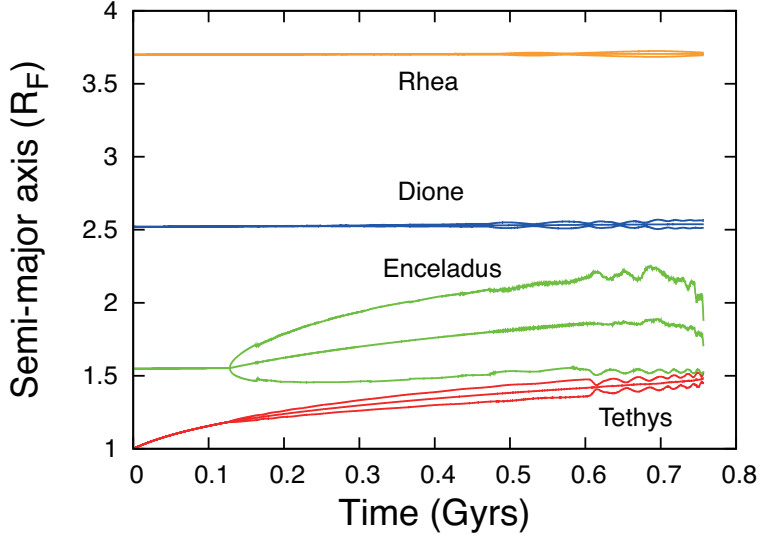}
  \caption{Evolution of semi-major axis, pericenter and apocenter of Enceladus, Tethys, Dione and Rhea in SET1A.  The speed-up factor $C$ for semi-major axis expansion and that for the eccentricity damping are $10^4$. Enceladus gets trapped in external 2:3 mean-motion resonance with Tethys. After the eccentricity of Enceladus is excited, hit-and-run collisions repeat a few dozens of times.}
   \label{fig: 4tai set1}
\end{figure}

Figure \ref{fig: 4tai set1} shows a typical result of orbital evolution of SET1A. 
In this case, the ring torque is not taken into account $(M_{\rm ring}=0)$. 
We adopt $Q_{\rm p} = 1\,700$ and $C = 10^4$.
Because interactions of Enceladus, Tethys and Dione are essential for
the orbital evolution in SET1A, we omit Mimas.

We start simulations when Tethys is formed at $a_{\rm T,0} \sim a_{\rm F}=1$.
Enceladus was already formed and has migrated to $a_{\rm E,0} \sim 1.5a_{\rm F}$ in agreement with Figure 1.  
Because Tethys is 5.7 times more massive (Table 1), it catches up with Enceladus (Eq.~\ref{eq:a_evol}) and Enceladus gets trapped in outer 2:3 mean-motion resonance  with Tethys' orbit at $t \simeq 0.13$ Gyrs.
Although the migration is accelerated by a factor of $C=10^4$, 
Enceladus gets trapped in the 1st order mean-motion resonance that it first encountered.
The trapping occurs if the convergent migration speed
is low enough (``adiabatic")  \citep{MurrayDermott1999}.
Thus, the trapping in 2:3 resonance would be robust in the realistic case with $C=1$.

As Enceladus migrates while being trapped in the resonance, the eccentricity of Enceladus ($e_{\rm{E}}$) secularly increases.
The analytical prediction for $j+1:j$ mean-motion resonance trapping
given by Eq.~(\ref{eq: e2sat_2}) in the Appendix A based on \citet*{Malhotra1995} shows
\begin{equation}
\frac{de_{\rm E}^2}{dt} \simeq \frac{1}{j+1} C 
\left(\frac{1}{a_{\rm T}} \frac{da_{\rm T}}{dt} - \frac{1}{a_{\rm E}} \frac{da_{\rm E}}{dt}\right)
\simeq C \frac{2.5}{j+1}\frac{k_{\rm 2p}}{Q_{\rm p}}
\frac{M_{\rm T}}{M_{\rm p}} \left(\frac{R_{\rm p}}{a_{\rm T}}\right)^5 \Omega_{\rm T},
\label{eq:e_inc_res}
\end{equation}
where we used Eq.~(\ref{eq:a_evol}) and assumed $M_{\rm T} \gg M_{\rm E}$.
The tidal $e$-damping is given by Eq.~(\ref{eq:e_damp}) as
\begin{eqnarray}
\frac{de_{\rm E}^2}{dt} = \frac{2 e_{\rm E}^2}{\tau_e}
= - C \frac{21 \Omega_{\rm E} M_{\rm p} R_{\rm E}^5}{M_{\rm E} a_{\rm E}^5}\frac{k_{\rm 2m}}{\qm}e _{\rm E}^2.
\end{eqnarray}
Equilibrating this excitation and damping with
$a_{\rm E}/a_{\rm T} = (j+1/j)^{2/3}$, the asymptotic value of $e_{\rm E}$ is estimated as 
\begin{eqnarray}
\label{eq: e-excite}
e_{\rm E} & \sim & \left[\frac{1}{7(j+1)}\left(\frac{j+1}{j}\right)^{13/3}
\left(\frac{M_{\rm E}}{M_{\rm p}}\right)^{1/3}
 \frac{M_{\rm T}}{M_{\rm E}} \frac{Q_{\rm m}/k_{\rm 2m}}{Q_{\rm p}/k_{\rm 2p}}\right]^{1/2} \nonumber \\
 & \sim & 0.3 \left[\frac{(j+1/j)^{10/3}/j}{2}\right]^{1/2}
\left(\frac{M_{\rm E}/M_{\rm p}}{0.19 \times 10^{-6}}\right)^{1/6}
\left(\frac{M_{\rm T}/M_{\rm E}}{5.7}\right)^{1/2}
\left(\frac{k_{\rm 2m}/Q_{\rm m}}{10^{-5}}\right)^{-1/2}
\left(\frac{k_{\rm 2p}/Q_{\rm p}}{10^{-4}}\right)^{1/2}.
\label{eq:e_asym}
\end{eqnarray}
 Although this analytical estimate includes uncertainty for such high $e$,
 the numerical simulation in Fig.~\ref{fig: 4tai set1} actually shows that
 the eccentricity of Enceladus secularly increases toward a high value,
until $e_{\rm E}$ and $e_{\rm T}$ become $\sim 0.18$ and $\sim 0.04$, respectively,
and orbital crossing starts between Enceladus and Tethys at $\sim 0.6$ Gyr. 
Note that the speed-up factor $C$ cancels out in this analytical estimate, 
suggesting that the equilibrium eccentricity would be similar in a real system with $C=1$.
After that, Tethys and Enceladus 
repeat a few tens of hit-and-run collisions, because the collision velocity is 
excited by the resonant secular perturbations and is significantly larger than $v_{\rm{esc}}$. 

We performed 50 runs in SET1 (A and B)
and a similar evolution was found in all cases except one run 
in which Enceladus was scattered to the inside of Tethys' orbit without collisions.
Initial conditions of semi-major axis of Tethys, Dione and Rhea, speed-up factor, mass of the ring, inclination of Tethys $i_{\rm T}$ and number of simulations are listed on Table 2.
In most runs, we adopted the same speed-up factor for tides of the moons and the planet.
In some runs, we adopted different values for 
the moons ($C_{\rm m}$) and the planet ($C_{\rm p}$).
The varies between the runs are only initial orbital angle of each moon.
Even in the inwardly scattered case, Tethys and Enceladus
collide many times after the inward scattering of Enceladus and before they become isolated.
We also performed runs with non-zero energy dissipation
at hit-and-run collisions.
In those cases, they still repeat collisions and eventually they merge
because the collision velocity becomes smaller as the collisions repeat. 

In some runs in SET1A, we include the ring torque in the initial condition of $a_{\rm E,0} > a_{\rm T,0}$.
We found that the results are similar. 
Therefore, we conclude that runs in SET1A
inevitably end up merging or disrupting Enceladus
and the current orbital configuration of the mid-sized moons is never reproduced.

In SET1B, Enceladus and Tethys migrate together in a horseshoe orbit
in the early phase.
We performed 5 runs of this case and
the initial conditions of semi-major axis of Tethys, Dione and Rhea, speed-up factor $C$, 
mass of the ring and number of simulations are listed on Table 2.
However, Enceladus and Tethys eventually start orbit crossing and there are repetitive hit-and-run collisions
as in SET1A runs.
Therefore, SET1B does not reproduce the current orbital configuration either.

\subsection{SET2: Enceladus posterior to Tethys}

In SET2, we start from the initial conditions with $a_{\rm T,0} > a_{\rm E,0}$.
We performed 30 runs in this set:
20 runs considering only semi-major axis expansion by the ring torque 
(SET2A: subsection 3.2.1 and 3.2.2) and 10 runs considering both semi-major axis expansion and eccentricity excitation by the ring torque (SET2B: subsection 3.2.3).

The current orbital separation between Enceladus and Tethys  
is slightly smaller than their 3:2 resonance.
As Fig.~\ref{fig:tide_evol_ring} shows, if the ring torque
can transfer Enceladus to the orbit inside the 3:2 resonance with Tethys,
the current orbital configurations may be reproduced.
Even if the ring torque sends Enceladus to the orbit beyond
4:3, 5:4, or a higher-$j$ resonance with Tethys, 
Enceladus is not trapped at these resonances, because 
the Tethys-Enceladus migration eventually becomes divergent at $a_{\rm E}>1.59$. 
Then Enceladus divergently passes the closer resonances to Tethys and approaches the 3:2 resonance again.
From Eq.~(\ref{eq:a_evol}), in the regions where the ring torque is not effective,
the ratio of tidal orbital expansion rate of Tethys to that of Enceladus is described as follows
\begin{equation}
\label{eq: e_t_migration}
\frac{(1/a_{\rm T})(da_{\rm T}/dt)}{(1/a_{\rm E})(da_{\rm E}/dt)} = \frac{M_{\rm T}}{M_{\rm E}}\left(\frac{a_{\rm T}}{a_{\rm E}}\right)^{-6.5}.
\end{equation}
Because $M_{\rm{T}}/M_{\rm{E}} \simeq 5.7$, this ratio is $> 1$ (divergent migration) for $a_{\rm{T}}/a_{\rm{E}} < 1.3$. 
Because the 3:2 resonance corresponds to $a_{\rm{T}}/a_{\rm{E}} \simeq 1.3$, the Tethys-Enceladus pair would not be trapped at resonances deeper (smaller $a_{\rm{T}}/a_{\rm{E}}$) than 3:2 resonance.
Therefore, one of the key points in SET2 is whether the ring torque
can send Enceladus to  an orbit closer to Tethys's orbit than the 3:2 resonance with Tethys.

\subsubsection{SET2A: Results with orbital expansion by ring torque}

As shown in Eq.~(\ref{eq:ring_dominate}), 
the ring torque dominates over the planetary tidal torque
for the orbital expansion near the ring.
Because the rate of orbital expansion by the ring torque is
proportional to $(\Delta a)^{-3}$, with $\Delta a$
the distance from the ring edge (Eq.~\ref{eq:ring_a}),
the expansion proceeds very rapidly near the ring edge and   
the Enceladus-Tethys migration is always initially convergent
in the setting of SET2 ($a_{\rm T,0} > a_{\rm E,0}$).   
From Eq.~(\ref{eq:ring_a}), in the ring torque dominated region,
\begin{equation}
\label{eq: e_t_migration}
\frac{(1/a_{\rm T})(da_{\rm T}/dt)}{(1/a_{\rm E})(da_{\rm E}/dt)} = \frac{M_{\rm T}}{M_{\rm E}}\left(\frac{a_{\rm T}}{a_{\rm E}}\right)^{1.5}
\left(\frac{\Delta a_{\rm T}}{\Delta a_{\rm E}}\right)^{-3}.
\end{equation}
Because $M_{\rm T}/M_{\rm E} \simeq 5.7$, the migration becomes divergent
when the Enceladus-Tethys pair migrates outward and $\Delta a_{\rm T}/\Delta a_{\rm E}$ decreases to be $\la 2$.

\begin{figure}[htbp]
  \centering
  \includegraphics[width=100mm, angle = 0]{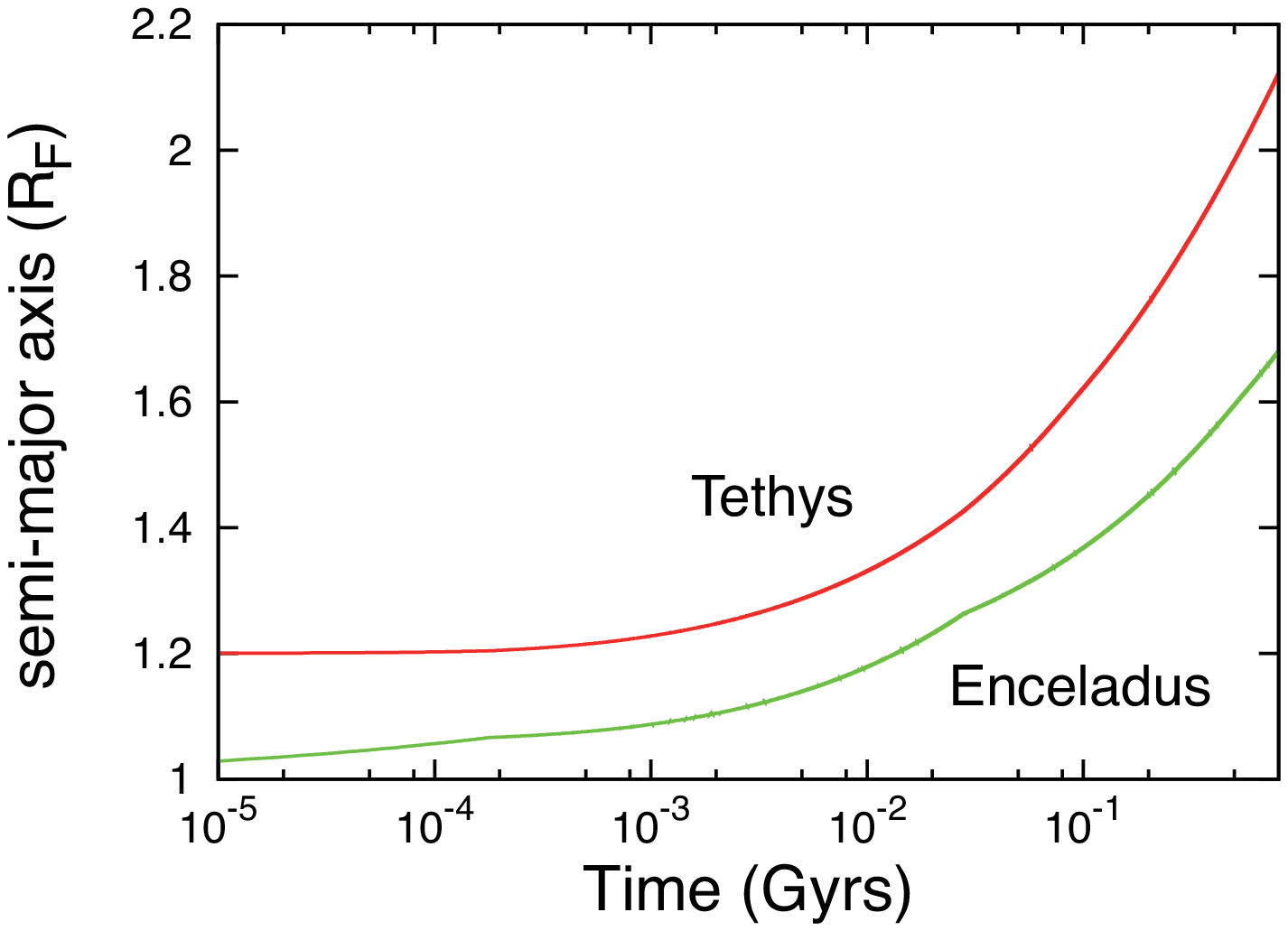}
  \includegraphics[width=100mm, angle = 0]{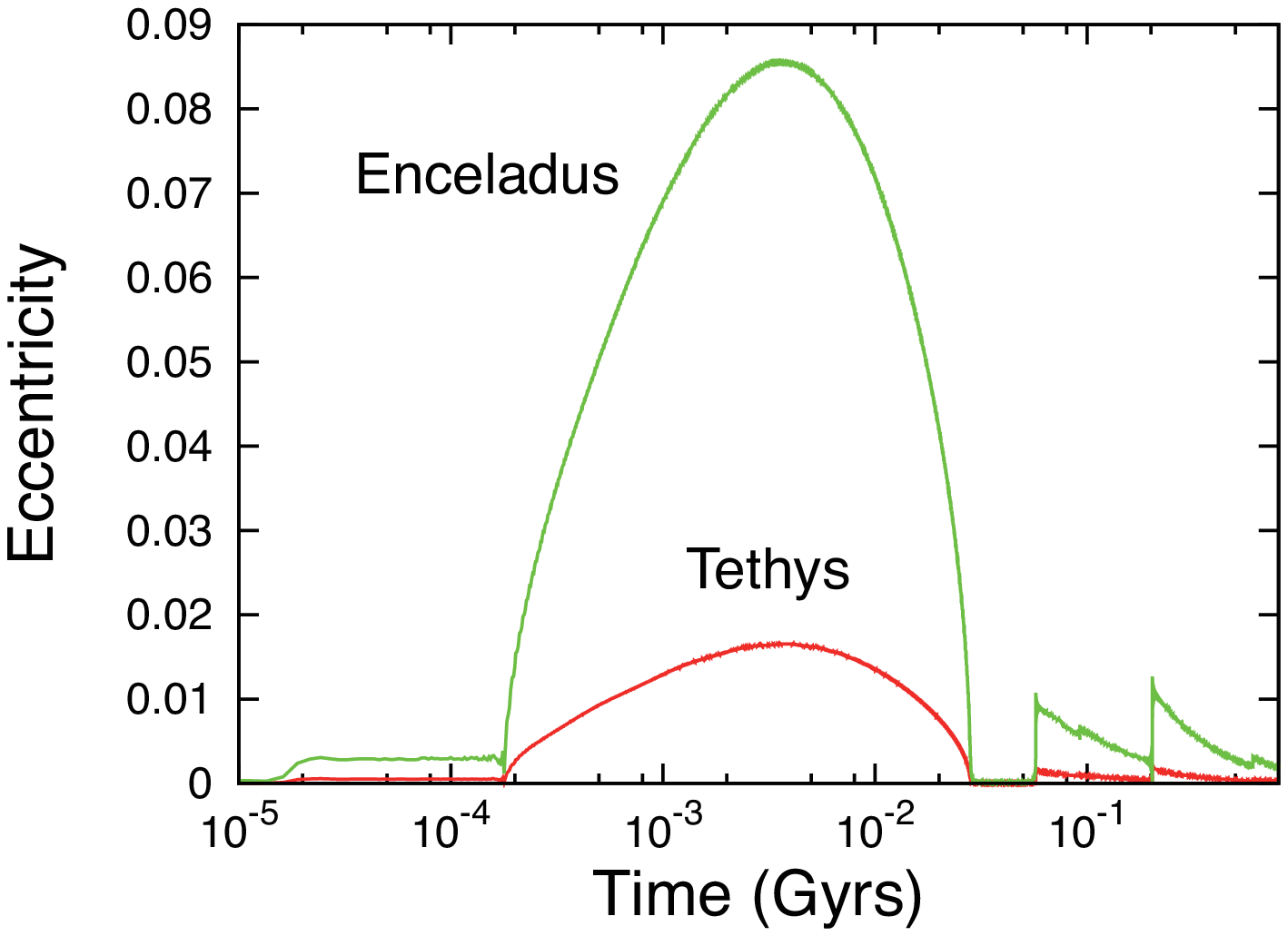}
  \caption{Evolution of semi-major axis of Enceladus (green) and Tethys (red). To focus on the initial part ($< 0.1$ Gyr), we took x-axis as logarithm scale. Enceladus is swept out from the ring soon after Tethys. Enceladus is captured in a 6:5 mean-motion resonance with Tethys and they migrate outward together until 0.03 Gyrs. }
  \label{fig: 2tai_t_e}
\end{figure}

In order to highlight the resonant interaction between Enceladus and Tethys, 
Figure~\ref{fig: 2tai_t_e} shows the orbital evolution in the case of
only Enceladus and Tethys with the effect of orbital expansion by the ring torque. 
Enceladus gets trapped in 6:5 resonance at $ t \simeq 2.0\times10^{-4}$ {\bf Gyrs}.
This is the 1st-order mean-motion resonance that Enceladus meets in the first place during its orbital evolution.
In this run, $C = 10^4$.
The probability of resonance trapping is higher for slower convergence of the migration.
In the real system with $C=1$, where migration is much slower,
Enceladus should also get trapped in the 1st-order mean-motion resonance that Enceladus meets at the first place. 
Enceladus' eccentricity increases after the resonant trapping, according to Eq.~(\ref{eq:e_inc_res}).

At $ t \sim 3\times10^{-2}$ Gyrs, $\Delta a_{\rm T}/\Delta a_{\rm E}$ becomes $\simeq 2$ and
the migration becomes divergent. 
Enceladus leaves the 6:5 resonance with Tethys
and its eccentricity decays. 
At $t=$ 0.06, 0.20 and 0.60, the Enceladus-Tethys pair pass through 
the 4:3, 7:5 and 3:2 mean-motion resonances, respectively, and their eccentricities are excited.
Because Tethys is 5.7 times more massive than Enceladus, the excited eccentricity is much larger for Enceladus than for Tethys. 
The maximum eccentricity is $\sim 0.08$ for Enceladus
and $\sim 0.02$ for Tethys in this run.
Enceladus can store much more heat in its interior than Tethys (see section 4).

Thus, SET2A conditions have the potential to reproduce 
the current orbital configurations of the mid-sized moons and
to account for the high thermal activity of Enceladus.
We performed 4 runs of this simulation and
the initial conditions of semi-major axis of Tethys, mass of the ring
and number of simulations are listed on Table 2.
However, runs adding Dione manifest a new problem, as shown below. 

\subsubsection{SET2A: Interaction with Dione}
 
Because Dione is more massive than Tethys/Enceladus and
Fig.~\ref{fig:tide_evol_ring} suggests that Dione may not have undergone
orbit crossing with other moons, we set Dione
at the time of birth of Enceladus at the location predicted by the backward integration
in Fig.~\ref{fig:tide_evol_ring}.
For this setting, we performed 16 runs.
The initial conditions of semi-major axis of Tethys and Dione, speed-up factor $C$, mass of the ring
and number of simulations are listed on Table 2.

\begin{figure}[htbp]
  \centering
 \includegraphics[width=70mm, angle = -90]{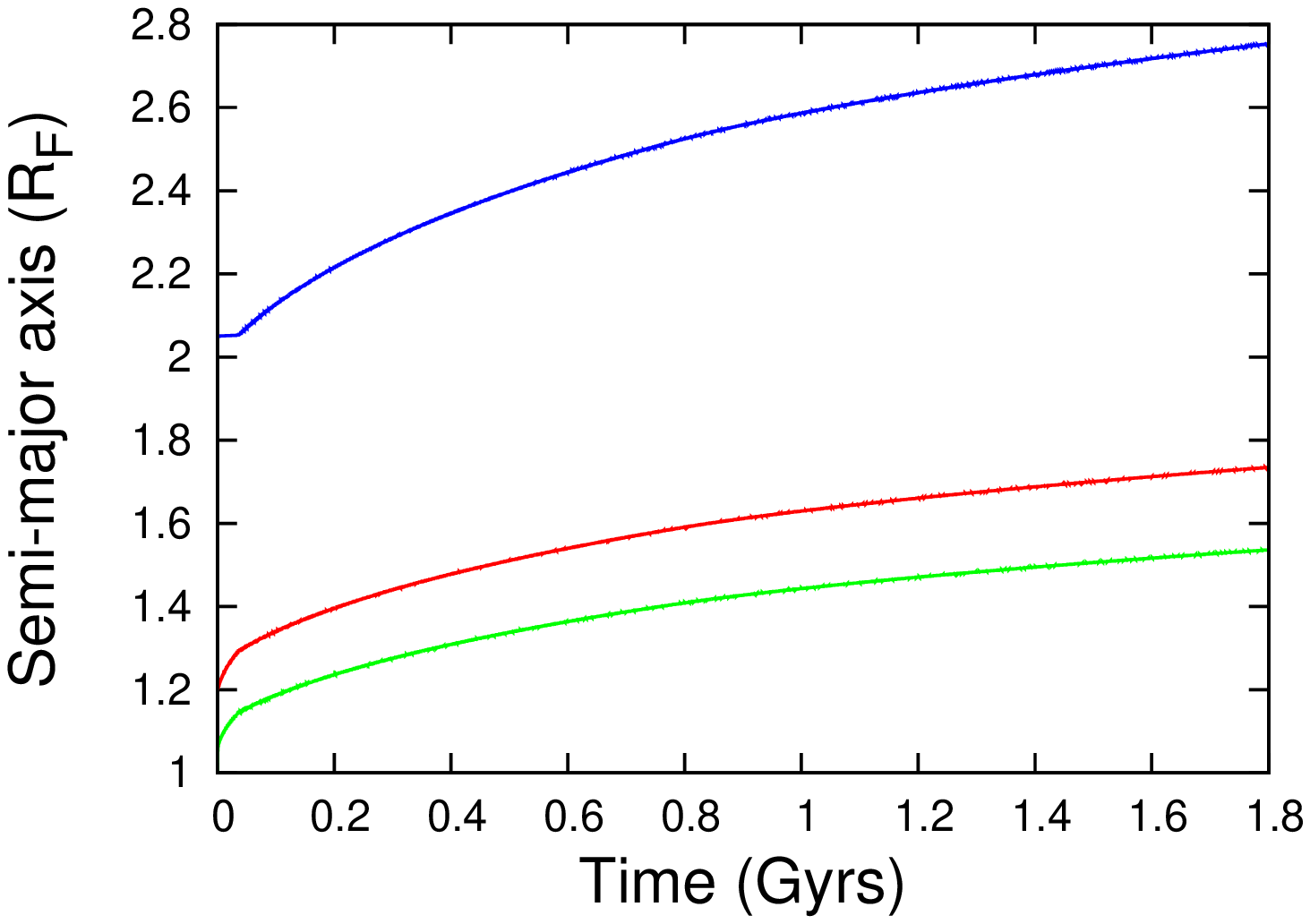}
 \includegraphics[width=70mm, angle = -90]{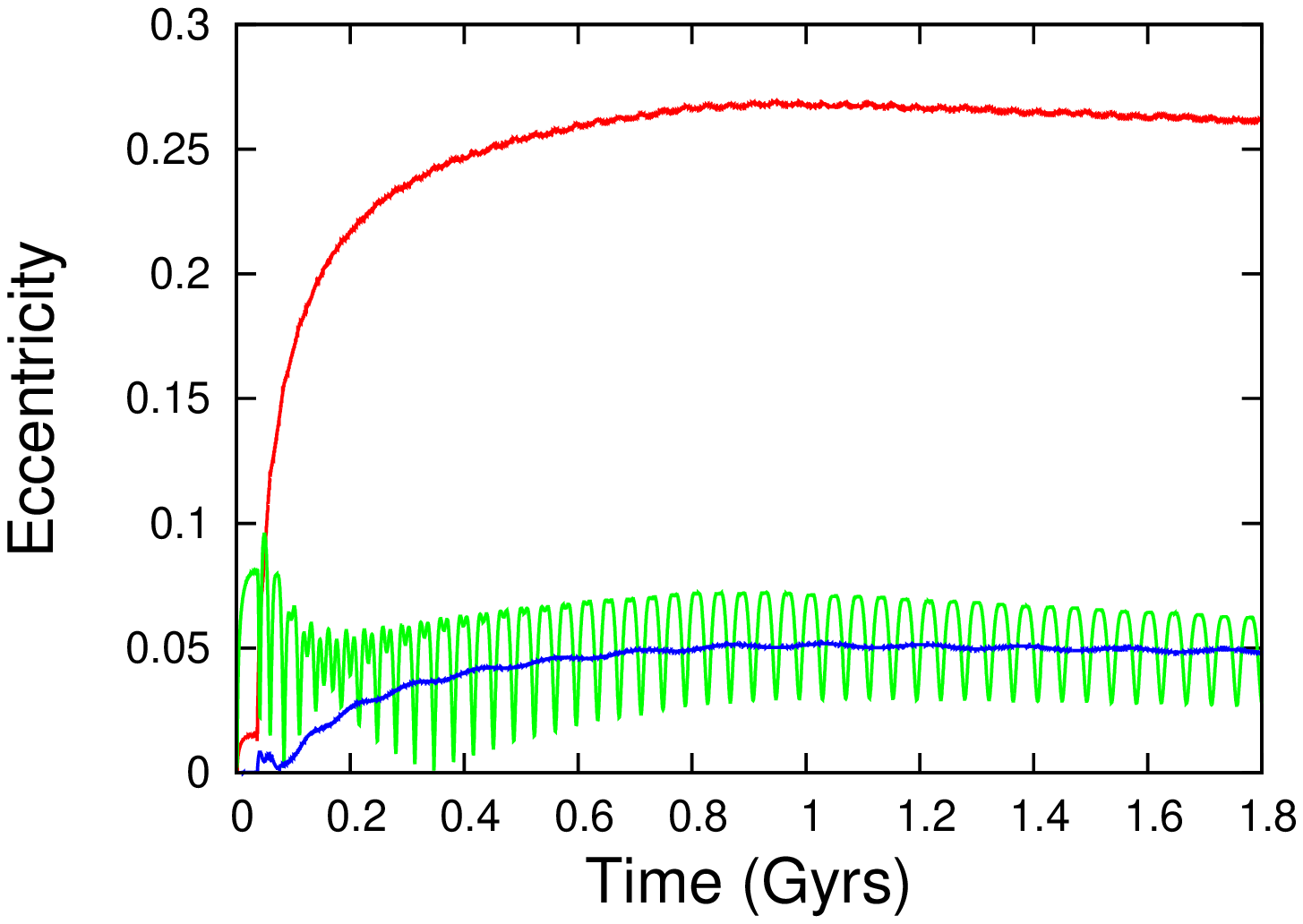}
  \caption{Evolution of semi-major axis of Enceladus (green), Tethys (red), and Dione (blue). Enceladus is swept out from the ring soon after Tethys. Enceladus is captured in a 6:5 mean-motion resonance with Tethys (same as the case without Dione). Tethys is captured in a 2:1 resonance with Dione. The resonances are kept to the end.}
  \label{fig: 3tai_t_e_d}
\end{figure}

Figure~\ref{fig: 3tai_t_e_d} shows the result of the run in which
Dione is added to the initial conditions of the run in Fig.~\ref{fig: 2tai_t_e}.
The early orbital evolutions of Enceladus and Tethys are the same as
the case in Fig.~\ref{fig: 2tai_t_e}:
Enceladus is captured in a 6:5 mean-motion resonance with Tethys and its eccentricity is secularly increased. 
However, in this case, Tethys rapidly gets trapped in a 2:1 resonance with Dione.
The Tethys-Dione migration is always convergent because Dione is only 1.8 times
more massive than Tethys, while Dione has sufficiently larger $a$ than Tethys.
Since Dione slows down Tethys' migration through the resonance,
the migration of Tethys and Enceladus remains convergent.
As a result, the multi-resonants state of 6:5 for Enceladus-Tethys and 2:1 for
Tethys-Dione is established and this configuration is stable until the end of the simulation. 
Due to the resonant migration, the eccentricities of the moons, 
and that of Tethys in particular, are secularly increased to values that are extremely high in comparison to the values at present.
Currently, orbital separation between Tethys and Dione is smaller than
3:2 resonance, which is inconsistent with the trapping in 2:1 resonance 
obtained by the simulation.
The other 9 runs show similar results.
Therefore, how to break up the Tethys-Dione's mean-motion resonance is a critical issue.  
 
 \subsubsection{SET2B: Results with both orbital expansion and eccentricity excitation by ring torque}
 \label{subsec:both}
 
For the resonant capture, 
adiabatic convergent migration is required.
In addition to that, for the capture, 
the eccentricity of a moon that encounters a mean-motion resonance 
with a more massive moon of mass $M_{\rm m}$ must be smaller than a critical value 
($ M_{\rm m}/M_{\rm p} \sim 2 \times 10^{-6}$ for Dione), 
given by \citep{Malhotra1993}
 \begin{equation}
\label{eq: e_crit}
e_{\rm{crit}} \simeq 1.58\left[ \frac{j}{(j+1)^2} \frac{\mm}{M_{\rm p}} \right]^{1/3}
\sim 0.01 \left[\frac{j/(j+1)^2}{0.2}\right]^{1/3}\left(\frac{\mm/M_{\rm p}}{10^{-6}}\right)^{1/3},
\end{equation}
For $e > e_{\rm{crit}}$, the capture probability for the $j+1:j$ resonance abruptly decays. 

We simulate the orbital evolution of the moons by varying the initial eccentricity of Enceladus $e_{\rm E,0}$ between 0.01 and 0.03,
while the other moons' eccentricities are set to 0.
We performed 11 runs of simulations in this set and
the initial conditions of semi-major axis of Tethys and Dione, dissipation factor of Saturn $\qp$, mass of the ring
and number of runs are listed on Table 2.

\begin{figure}[htbp]
 \begin{minipage}{0.5\hsize}
  \begin{center}
\includegraphics[width=50mm, angle = -90]{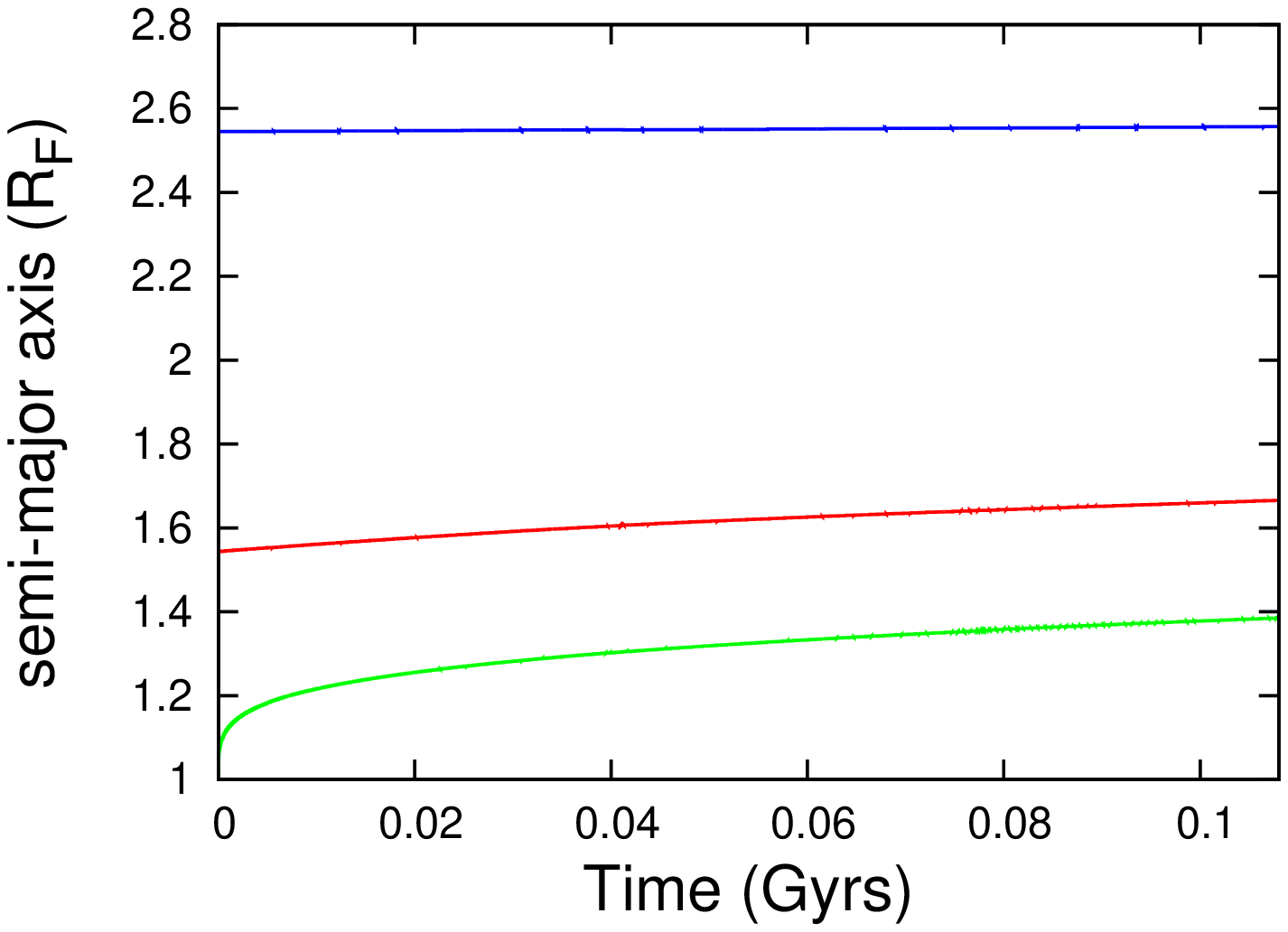}
  \end{center}
 \end{minipage}
 \begin{minipage}{0.5\hsize}
  \begin{center}
\includegraphics[width=50mm, angle = -90]{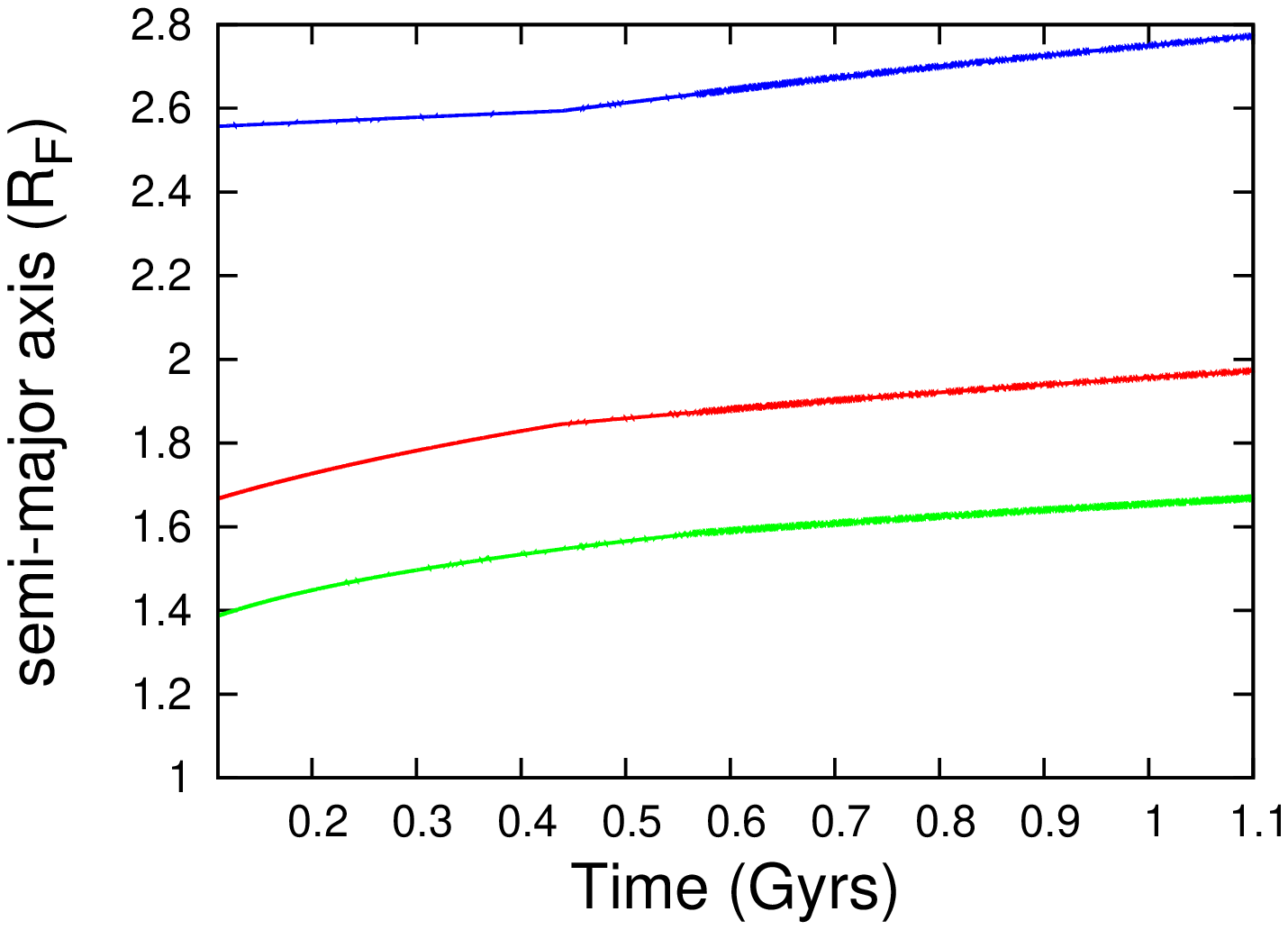}
  \end{center}
 \end{minipage}\\
  \begin{minipage}{0.5\hsize}
  \begin{center}
   \includegraphics[width=50mm, angle = -90]{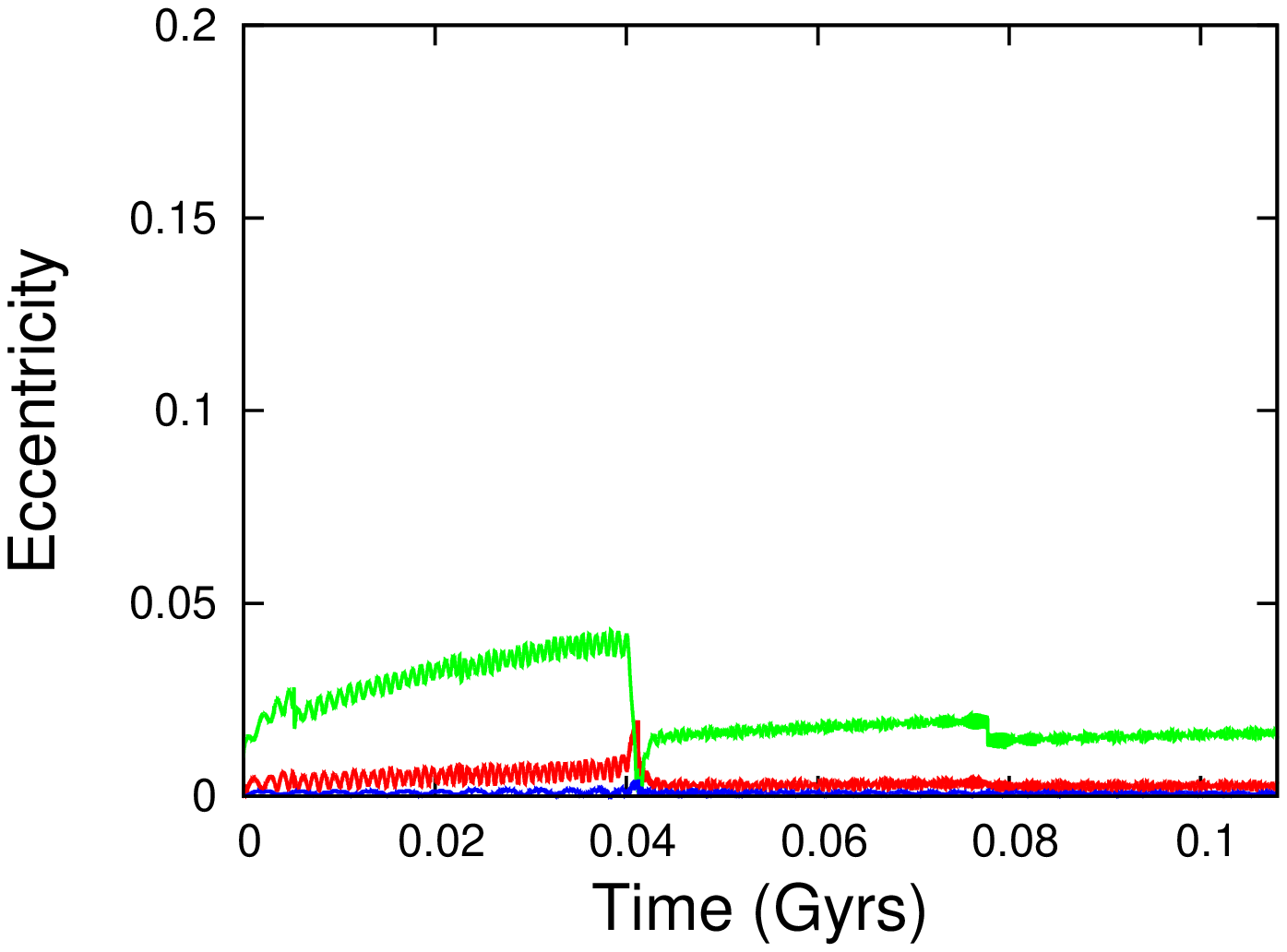}
  \end{center}
 \end{minipage}
  \begin{minipage}{0.5\hsize}
  \begin{center}
   \includegraphics[width=50mm, angle = -90]{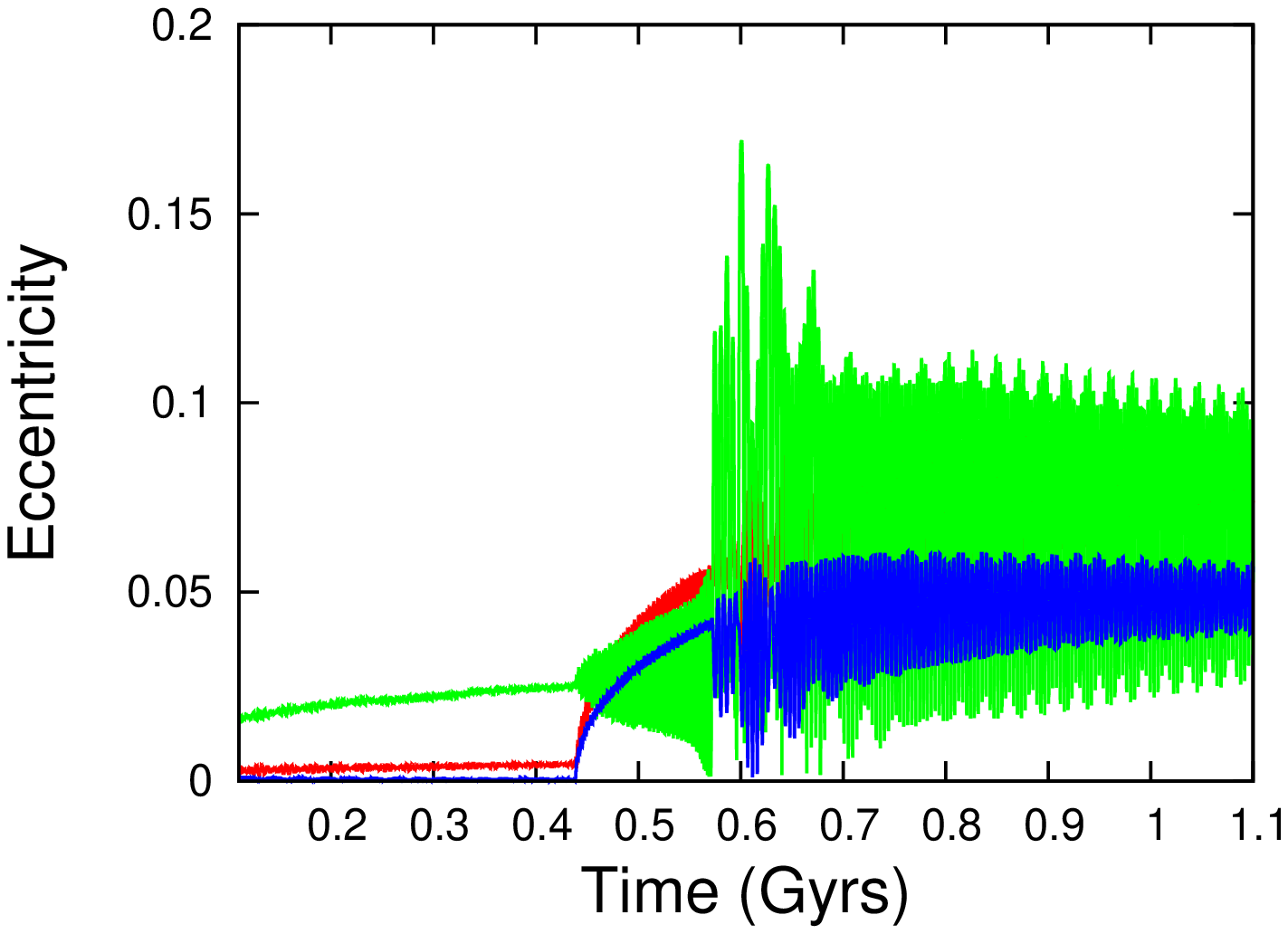}
  \end{center}
 \end{minipage}
 \caption{The evolution of semi-major axis (upper) and eccentricity (lower) of Enceladus (green), Tethys (red), and Dione (blue) with $e_{\rm E,0}=0.01$.
In the left panels, the speed-up factor of $C = 10^3$ is used.
Tethys passes 2:1 resonance with Dione at $t \simeq 0.04$ Gyrs.
The results in the right panels start from the end state of the evolution in the left panels and use $C=10^4$. Note that the oscillation patterns of the eccentricities are modified by the change in $C$. Tethys is captured in a 5:3 resonance with Dione at $t \simeq 0.45$ Gyrs and immediately afterward, Enceladus is trapped into a 9:7 resonance with Tethys (the right panels).
}
  \label{fig:q4000_ver1}
\end{figure}

In Fig.~\ref{fig:q4000_ver1}, we include the eccentricity excitation due to the ring torque with $M_{\rm ring} = 4 M_{\rm E}$
and $e_{\rm E,0} = 0.01$ for Enceladus. 
In the beginning of Fig.~\ref{fig:q4000_ver1} (on the left of this figure), we accelerated the tidal orbital evolution with $C=10^3$, 
and adopted $\km/\qm$ as $10^{-5} $.
Equation~(\ref{eq:ring_e_dominate}) predicts that
the eccentricity increases until $a$ reaches $(\Delta a)_{{\rm crit},e} \simeq 0.3$ 
for Enceladus, which is consistent with the numerical result.
In the numerical result, $e_{\rm E}$ is already excited up to $\sim 0.04$
at the timing of Tethys-Dione 2:1 resonance passing.
Because $e_{\rm E}$ is well excited beyond $e_{\rm crit}$,
Enceladus passes through 3:2, 4:3 and closer to 1st order resonances with Tethys,
although the Enceladus-Tethys migration is convergent until Enceladus
reaches $a_{\rm E} \sim 1.59$ and the ring torque decays.

\begin{figure}[htbp]
 \includegraphics[width=110mm, angle = -90]{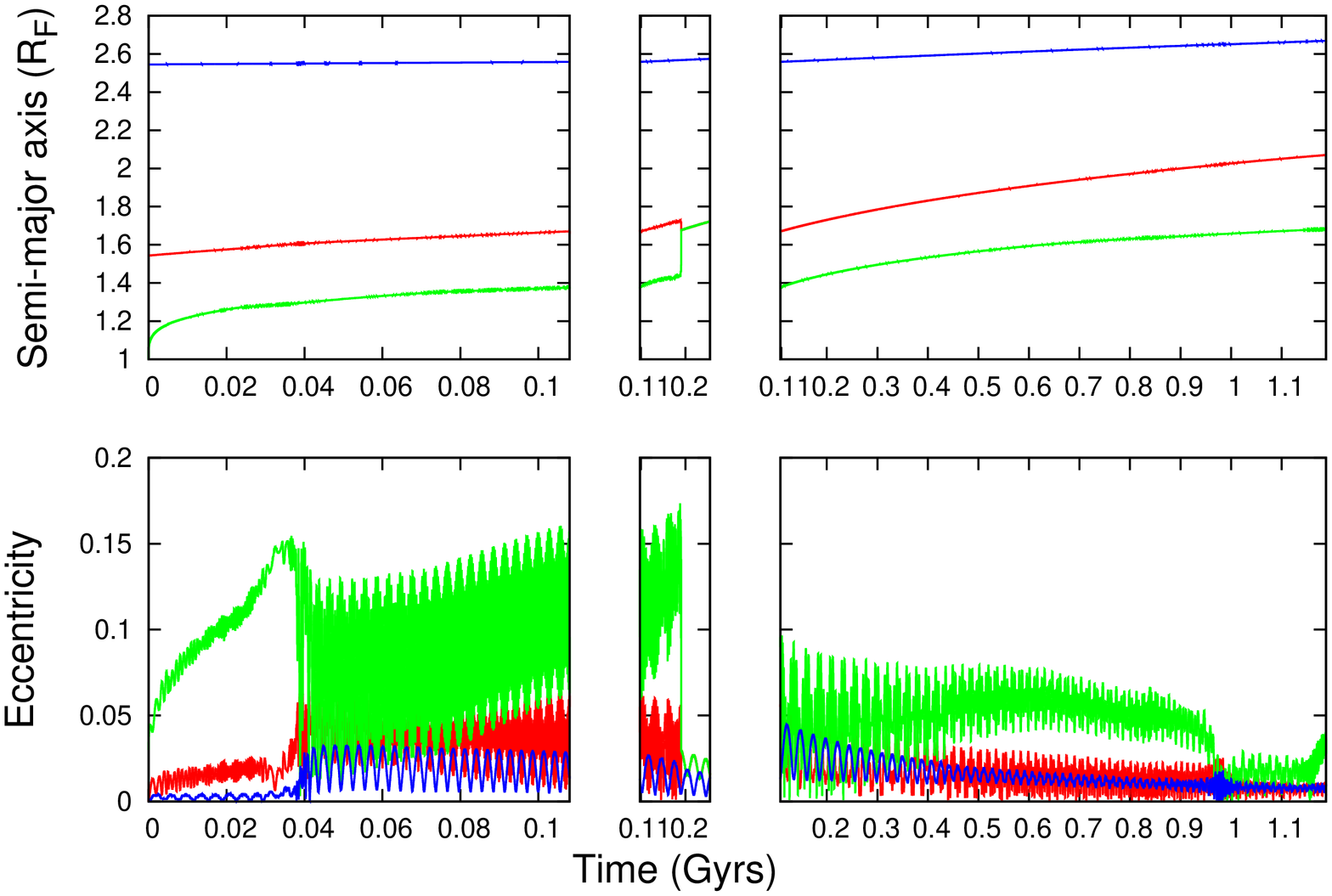}
 \caption{Same as Fig.~\ref{fig:q4000_ver1}, except $e_{\rm E,0} = 0.03$.
In the left panels, Tethys passes 2:1 resonance with Dione at $t \simeq 0.04$ Gyrs, as in the results of Fig.~\ref{fig:q4000_ver1}. 
The middle panels start from the end state of the evolution in the left panels and use $C=10^4$. 
In this case, Tethys and Enceladus collide with each other at $\sim 0.18$ Gyr.
However, when the eccentricity of Enceladus becomes smaller, 
they never collide and it can reproduce the orbital configuration (the right panels).
At the end of the right panels ($\sim  1.1$ Gyr), Enceladus and Dione get trapped in 2:1 resonance, which is the current state.}
  \label{fig:q4000_ver8}
\end{figure}

Secular perturbation from Enceladus with the relatively high eccentricity
enhances Tethys' eccentricity up to $e_{\rm T} \sim 0.01 \sim e_{\rm crit}$.
Tethys encounters a 2:1 resonance with Dione at $t \simeq 0.04$ Gyrs.
At the resonance passage, a relatively large amount of angular momentum is
exchanged between Tethys and Dione and also between Tethys and Enceladus.
Owing to $e_{\rm T} \sim e_{\rm crit}$,
Tethys successfully avoided getting trapped in a 2:1 resonance with Dione.

In Fig. \ref{fig:q4000_ver1}, the tidal orbital evolution is accelerated from $C=10^3$ (the left panels) to 
$C=10^4$ (the right panels) to follow the whole orbital evolution.
After $a_{\rm E}$ exceeds 1.59 at 0.6 Gyr,
the Enceladus-Tethys migration becomes divergent and
it becomes impossible for them to be trapped in 1st-order resonances between them.
The calculation of the following evolution shows that the final 
$a_{\rm T}$ and $a_{\rm D}$ are consistent with their current semi-major axes ($a_{\rm T}= 2.11$ and $a_{\rm D}= 2.70$). 
However, both Tethys' and Dione's eccentricity ($e_{\rm T}$ and $e_{\rm D}$, respectively) 
are enhanced too much by the resonance in the simulation. 
Indeed, the Tethys-Dione pair gets trapped in a 5:3 resonance
at $t \simeq 0.45$ Gyrs and the resonance configuration remains stable, 
while the current Tethys-Dione separation
is slightly smaller than the 3:2 resonance.
Immediately after that, Enceladus gets trapped in a 2nd-order 9:7 mean-motion resonance with Tethys.
These resonances are not consistent with the current orbit.

In this case, a mechanism is needed to kick the Tethys-Dione pair
out of their resonance, such as 
the impact that created the Odysseus crater on Tethys \citep{Zhang2012}.
If we consider the growth of the moons by merging as they migrate outward,
it is, in principle, possible that the moons avoid the resonance capture.
However, N-body simulations of collisional growth of
the mid-sized moons by \citet{Salmon2017} 
showed that capturing at Tethys-Dione's 2:1 and 3:2 resonances is still a great difficulty
(their Figs. 11 and 12),
although they did not discuss this issue. 

In Fig.~\ref{fig:q4000_ver8}, we set $e_{\rm E,0} = 0.03$ and other parameters were the same as 
those in the results of Fig.~\ref{fig:q4000_ver1}. 
With the three times higher $e_{\rm E,0}$,
the eccentricity of Enceladus increases up to $e_{\rm E}\sim 0.15$.
In addition, Tethys' eccentricity is also enhanced, because the effect of 
the secular perturbation is proportional to the perturber's eccentricity.
At $t\simeq 0.04$ Gyrs, as $e_{\rm T}$ is enhanced well beyond $e_{\rm crit}$,
Tethys passes through the Tethys-Dione 2:1 resonance 
and angular momentum is transported between the moons (the left panels). 
The eccentricities of all the moons are excited and oscillate  substantially from $t\simeq 0.04$ Gyrs to the point of $a_{\rm E} \simeq 1.59$,
where the excitation of eccentricity and semi-major axis by the ring torque 
for Enceladus decays. 

In the middle panels, the following orbital evolution is calculated with $C = 10^4$.
As $e_{\rm D}$ is excited to $\sim 0.1-0.15$, 
Enceladus eventually collides with Tethys at $t = 0.18$ Gyrs.
We performed 11 runs with orbital expansion and
eccentricity excitation by the ring torque.
In 2 runs with $e_{\rm E,0} \la 0.03$, the orbital evolution is similar to Fig.~\ref{fig:q4000_ver1} and 
Tethys is captured in the Tethys-Dione 3:2 resonance.
In other 8 runs with $e_{\rm E,0} \ga 0.04$, $e_{\rm E}$ is excited so much that Enceladus undergoes
collisions or close scattering with Tethys. 
Therefore, it is difficult for Tethys to successfully pass the 3:2 resonance with Dione 
and simultaneously avoid a collision with Enceladus,
as long as only Enceladus, Tethys and Dione are integrated.

We consider a hypothetical case in which 
the eccentricity of Enceladus is smaller than in the middle panels.
In the right panels in Fig.~\ref{fig:q4000_ver8},
we artificially decreased $e_{\rm E}$ by a factor of 2 from the final state of 
the left panels and calculated the following orbital evolution.
Although we do not specify the cause of the decrease, 
the ring mass decrease due to the birth of Mimas could be responsible for it.
In this case, Enceladus and Tethys never collide.
But, the eccentricity of Tethys excited by secular perturbations from Enceladus
is still large enough
for Tethys to avoid getting trapped in the 3:2 resonance with Dione.
After passing the 3:2 resonance, both $e_{\rm T}$ and $e_{\rm D}$ are damped.
Enceladus is eventually trapped into the 2:1 resonance 
with Dione, which is the current resonance relation, because both $e_{\rm D}$ and $e_{\rm E}$
are damped below $e_{\rm crit}$.
After the trapping, 
it is predicted that $e_{\rm E}$ increases to an equilibrium value, $e_{\rm E} \sim 0.04$ (Appendix A),
which is 10 times larger than the current value.
Hence, Enceladus must have recently become trapped in the 2:1 resonance with Dione.

So far, we have neglected Mimas because it is the smallest mid-sized moon.
However, if the ring is still massive enough after Mimas' formation,
the torque from the ring can be transported to Enceladus, Tethys and Dione
through Mimas, because Mimas is currently located within a 2:1 resonance with the ring edge 
and should have suffered the ring torque throughout its entire orbital evolution.
Figure~\ref{fig: kaiseki} suggests that
Mimas-Enceladus encounters their 3:2 resonances
at a similar time as the trapping of the Tethys-Dione pair at their 3:2 resonance.
It is very likely that the ring torque pumps up Mimas' eccentricity to a value 
larger than $e_{\rm crit}$ and the Mimas-Enceladus pair avoids
the trapping at their 3:2 resonance.
However, interactions among the four moons with the ring torque are complicated.
The ring mass should also change at the formation of Mimas.
Because these investigations require much more parameter surveys,  
we leave them to future study.

\begin{table}
\begin{center}
\caption{Initial conditions}
\scalebox{0.9}{
\begin{tabular}{c|cccccccccc} \hline
SET  & $a_{\rm E,0}$  ($a_{\rm F}$) & $a_{\rm T,0}$  ($a_{\rm F}$) & $a_{\rm D,0}$  ($a_{\rm F}$) & $a_{\rm R,0}$  ($a_{\rm F}$) & $C_{\rm m}$ & $C_{\rm p}$ & $i_{\rm T} \, (\rm degree) $ & $M_{\rm ring}$ & $\qp$ & runs \\ \hline \hline 
        & 1.55 & 1.0 & 2.5 & 3.7 & $10^4$ & $10^4$ & 0 & 0 & 1700 & 10\\
        & 1.55 & 1.0 & 2.5 & 3.7 & $10^3$ & $10^4$ & 0 & 0 & 1700 & 10 \\
        & 1.55 & 1.0 & 2.5 & 3.7& $10^4$ &  $10^5$ & 0 & 0 & 1700 & 5\\
1A    & 1.55 & 1.0 & 2.5 & 3.7 & $10^5$ &$10^5$ & 0 & 0 & 1700 & 5  \\
        & 1.55 & 1.0 & 2.5 & 3.7 &$10^5$ & $10^4$ & 0 & 0 & 1700 & 5\\
        & 1.6   & 1.0 & 2.5 & 3.7 &$10^5$ & $10^5$ & 1.8 & 0 & 1700 & 5 \\
        & 1.6   & 1.0 & 2.5 & 3.7 &$10^5$ & $10^5$ & 3 & 0 & 1700 & 5 \\  \hline \hline
        & 1.0 & 1.0 & 2.5  & 3.7 & $10^5$ & $10^5$ & 0 & 0 & 1700 & 1\\
1B    & 1.0 & 1.0 & 2.5  & 3.7 & $10^6$ & $10^6$ & 0 & 0 & 1700 & 2  \\
        & 1.0 & 1.0 & 2.5 & 3.7 & $10^4$ & $10^4$ & 0 & 0 & 1700 & 2 \\ \hline \hline
        & 1.0 & 1.2   & -- & -- & $10^4$ & $10^4$ & 0 & $2M_{\rm E}$ & 1700 &  2\\
        & 1.0 & 1.18 & -- & -- & $10^4$ & $10^4$ & 0 & $2M_{\rm E}$ & 1700 &  2\\  
        & 1.0 & 1.25 & 2.05 & -- & $10^4$ & $10^4$ & 0 & $2M_{\rm E}$ & 1700 & 1\\
        & 1.0 & 1.3  & 2.2 & -- & $10^4$ & $10^4$ & 0 & $2M_{\rm E}$ & 1700 & 1\\
        & 1.0 & 1.35 & 2.05 & -- & $10^4$ & $10^4$ & 0 & $2M_{\rm E}$& 1700 & 1\\
        & 1.0 & 1.22 & 1.35 & -- & $10^4$ & $10^4$ & 0 &$2M_{\rm E}$ & 1700 & 1 \\
        & 1.0 & 1.15 & 1.35 & -- & $10^4$ & $10^4$ & 0 & $2M_{\rm E}$ & 1700 & 1\\
2A    & 1.0 & 1.4 & 2.2 & -- & $10^4$ & $10^4$ & 0 & $2M_{\rm E}$ & 1700 & 1\\
        & 1.0 & 1.2 & 2.15 & -- & $10^4$ & $10^4$ & 0 & $2M_{\rm E}$  & 1700 & 1 \\
        & 1.0 & 1.2 & 1.55 & -- & $10^4$ & $10^4$ & 0 & $2M_{\rm E}$ & 1700 &  1\\
        & 1.0 & 1.2 & 1.75 & -- & $10^4$ & $10^4$ & 0 & $2M_{\rm E}$ & 1700 & 1 \\
        & 1.0 & 1.2 & 2.05 & -- & $10^4$ & $10^4$ & 0 & $2M_{\rm E}$ & 1700 & 3 \\
        & 1.0 & 1.2 & 1.45 & -- & $10^4$ & $10^4$ & 0 & $2M_{\rm E}$ & 1700 & 1 \\
        & 1.0 & 1.2 & 1.35 & -- & $10^4$ & $10^4$ & 0 & $2M_{\rm E}$ & 1700 & 1 \\
        & 1.0 & 1.2 & 2.2 & -- & $10^4$ & $10^4$ & 0 & $2M_{\rm E}$ & 1700 & 1 \\
        & 1.0 & 1.2 & 2.25 & -- & $10^4$ & $10^4$ & 0 & $2M_{\rm E}$ & 1700 & 1 \\ \hline  \hline
        & 1.0 & 1.44 & 2.54 & --  & $10^3$ & $10^3$ & 0 & $7M_{\rm E}$ & 2000 & 1\\
2B    & 1.0 & 1.5 & 2.54   & -- & $10^3$ & $10^3$ & 0 & $5M_{\rm E}$ & 3000 & 1\\
        & 1.0 & 1.54 & 2.54 & --  & $10^3$ & $10^3$ & 0 & $4M_{\rm E}$ & 4000 & 8 \\
        & 1.0 & 1.5 & 2.54   & -- & $10^3$ & $10^3$ & 0 & $3M_{\rm E}$ & 5000 & 1 \\ \hline
\end{tabular}
}
\end{center}
\label{tab:initial cond}
\end{table}

\section{Heat Flux}
\label{sec: heat estimation}

As we have shown, the moons would have undergone
a high eccentricity phase in the past during orbital evolution.
As we show below, the heat generated during the high eccentricity phase can be
stored in the interior and the current high heat flux can reflect the stored heat
(the current heat generation is not balanced with the surface heat flux). 
From the numerical simulations, here we calculate the stored heat energy
for each moon,
\begin{equation}
E \sim \int H dt,
\end{equation}
where $H$ is given by Eq.~(\ref{eq:HEnceladus}).

Although Enceladus would have a subsurface ocean at present, most parts of bulk Enceladus
would be in a solid phase.
For simplicity, we assume that conduction is a major heat transfer mechanism
in the interior of the mid-sized moons.
We estimate the conduction timescale very roughly.
The thermal conductivity and the specific heat capacity of solid ice
are 2 W/m\,K and 2000 J/kg, and
those of rock are 3 W/m\,K and 900 J/kg, respectively.
Assuming the densities of ice and rock are $ 1000\,{\rm kg\,m^{-3}}$ and 
$\sim 3000\,{\rm kg\,m^{-3}}$, we can estimate the volume fraction, $x$, of rock 
by bulk density of moons; $\rho\,({\rm g\,cm^{-3}}) \sim (1-x) + 3x = 1+2x$.  
Using the obtained $x$, 
the mean thermal conductivity and heat capacity are
$\lambda \sim [2(1-x) + 3x] \, {\rm W/m\,K}$ and 
$\rho c \sim \{(1+ 2x) \times 10^3 [2000(1-x)+900x]\} \, {\rm Jm^{-3}/K}$, respectively.
The thermal diffusion coefficient of Enceladus is then
\begin{equation} 
\kappa = \frac{\lambda}{\rho c} \sim \zeta \times 10^{-6} \, \textrm m^2/\textrm {s},
\end{equation}
where $\zeta = (2+x)/[(1+2x)(2-1.1x)] \sim 1$.
The thermal conduction timescale for a moon with a physical surface radius of $R$ is 
\begin{equation}
\tau_{\rm cond} \sim \frac{R^2}{3 \kappa} \sim 0.6 \zeta^{-1} \left(\frac{R}{250\,{\rm km}}\right)^2\, {\rm Gyrs}.
\end{equation}
Because $\tau_{\rm cond}$ may be longer than the age of Enceladus in the low $Q_{\rm p}$ model,
we simply assume that most of the heat energy generated during the 
high eccentricity phase is still stored in the interior.
Then, the total heat flux is given by
\begin{equation}
L \sim \frac{E}{\tau_{\rm cond}}.
\end{equation}
For $e_{\rm max}$ being a typical $e$ of the high $e$ phase,
$H_{\rm max}$ being the heat generation rate for $e_{\rm max}$,
and $\Delta t$ being the duration  of the high $e$ phase, 
$L \sim H_{\rm max} (\Delta t/\tau_{\rm cond})$.
From Eq.~(\ref{eq:HEnceladus}), for example, 
if $e_{\rm max} \sim 0.2, \,k_{2m}/Q_m \sim 10^{-4}$ and 
$(\Delta t/\tau_{\rm cond}) \sim 0.1$, then $L \sim 16 \,{\rm GW}$.
In other words, 
\begin{equation}
e  \sim 
0.2 \left( \frac{k_{\rm 2m}/Q_{\rm m}}{10^{-4}}\right)^{-1}\left( \frac{R_{\rm m}}{250{\rm km}}\right)^{-5}\left( \frac{a_{\rm m}}{2.4 \times 10^{5}{\rm km}}\right)^{-15/2} \left(\frac{\Delta t/\tau_{\rm cond}}{0.1}\right)^{-1}
\end{equation}
is need to reproduce the current heat flux.

We integrate $H$ with $t$ to obtain the energy $E$ from the data of the numerical simulations 
and the total heat flux $L$ for individual simulations are listed in Table \ref{tidal_heat_result}. 
In the case corresponding to Fig. 4, the tidal heat is consistent with the current observed value.
However, in the results in Figs. \ref{fig: 3tai_t_e_d} and \ref{fig:q4000_ver1}, 
Tethys is captured by a resonance with Dione and subsequently has higher tidal heat generation than Enceladus.
In the case of Fig.~\ref{fig:q4000_ver8}, tidal heat of Enceladus is not inconsistent with the observed value 
which is suggested by \citet{Howett2011,Spencer2013}.
While the heat flux of other moons is smaller than Enceladus, the heat flux of Tethys predicted by our simulation is not sufficiently small, which could be consistent with the geological features of Tethys \citep{Giese2007,Chen2008}.
The case of Fig. \ref{fig:q4000_ver8} may be the most preferable case not only in final orbital configurations but also in the heat flux.

\renewcommand{\arraystretch}{1.5}
\begin{table}
\begin{center}
\begin{tabular}{c|cccc} \hline
                    & Fig. 4 & Fig. 5                    & Fig. 6  & Fig. 7\\ \hline \hline 
Enceladus   & 19      & 3.3                        &  6.3     & 6.4 \\
Tethys          & 2.5     & $1.9 \times10^2$ & 12      &  3.1 \\
Dione           & --           & 0.19                   & 0.44    & 0.14 \\ \hline
\end{tabular}
\end{center}
\caption{Calculated heat flux of Enceladus, Tethys, and Dione based on our simulations of SET2. These values are shown in GW units.}
\label{tidal_heat_result}
\end{table}
\renewcommand{\arraystretch}{1}

\section{Conclusion and discussion}
\label{sec: conclusion}
Through N-body simulations, we have numerically investigated the orbital evolution of 
Saturn's mid-sized moons (mainly Dione, Tethys and Enceladus), under the influence of
Saturn's tidal force, tidal dissipation in the moons, and the torque exerted by its ring.
Our work was based on the model of the mid-sized moons having formed relatively recently 
from the spreading out of a massive ring, a theory that was proposed by \citet{Charnoz2011}.
We have performed 80 runs in total with various initial conditions
at two different settings.

If the ring torque is sufficiently weak, Enceladus must be formed prior to
Tethys and scattered inward across Tethys' orbit. 
In this set of runs (SET1), we found that Enceladus is always trapped 
in the outer 1st-order mean-motion resonance with Tethys 
due to the rapid migration of Tethys which is more massive than Enceladus.
Such resonance trapping is inevitable 
because of the tidal migration timescale of these moons.
The eccentricity of Enceladus is secularly increased by the resonant migration.
When the eccentricity becomes large enough, Enceladus undergoes 
collisions with Tethys until the end of simulation. 
Therefore, it is impossible for Enceladus to cross Tethys' orbit and become isolated from Tethys, and the current orbital configuration of Enceladus and Tethys is never reproduced.

If the ring mass is comparable to the mass of a forming moon,
which is a reasonable assumption, the torque from the ring is strong enough
that Enceladus can be formed after Tethys, and 
Enceladus need not cross Tethys' orbit.
We also performed many simulations with many different initial conditions 
in the setting in which Enceladus is initially located inside Tethys' orbit (SET2).
Because the ring torque is very strong near the outer edge of the ring,
Enceladus-Tethys migration is convergent as long as Enceladus' orbital radius
is smaller than the 2:1 resonance with the ring edge.
As a result, Enceladus is always trapped in the 1st-order mean-motion resonance
with Tethys that Enceladus meets in the first place.
After the trapping, Enceladus' eccentricity is secularly increased to
undergo repeated hit-and-run collisions as in SET1 runs.
However, for some range of initial conditions, the migration turns into
a divergent one before the hit-and-run collisions start. 
After that, Enceladus is never trapped at 1st-order resonances with Tethys and 
orbital eccentricity decays via the tidal damping, which results in 
the final orbits of Enceladus and Tethys being consistent with the current ones.

However, Tethys is trapped in the 2:1 or 3:2 mean-motion resonance with Dione 
in this set of simulations, because their migration is always convergent
and adiabatic.
It is very difficult to reproduce their current orbital separation closer than the 3:2 resonance relation between Dione and Tethys.

Fuller et al. (2016) considered time-dependent $\qp$, in which
$\qp \ga 10000$ in the initial phase and it decreases down to $\qp \sim O(1000)$
after resonant locking between the orbital frequency and Saturn's oscillation mode.
In this case, moons can be formed in an extended circumplanetary disk
and orbit crossing of the moons does not occur.
Even in this model, the migration between Dione and Tethys is usually
convergent and adiabatic and it is difficult for them to avoid becoming trapped in the 3:2 resonance.

We found that, if eccentricity excitation by the ring torque is effective,
it has the potential to solve the problem.  
This excitation is effective only in the regions close to the outer edge of the ring.
Owing to the excited eccentricity, Enceladus can easily pass through the resonances with Tethys.
Enceladus' eccentricity could be mostly enhanced by the ring torque, but not by the 
resonant perturbations.
The modest eccentricity of Tethys raised by the secular perturbation from Enceladus
with relatively high eccentricity breaks the 2:1 resonance between Tethys and Dione. 
In part of the range of Enceladus' eccentricity, Tethys can also pass the 3:2 resonance with Dione.
As the distance between Enceladus and the ring increases,
tidal eccentricity damping dominates over the excitation by ring torque,
and the moons' eccentricities decay,
which enables the Enceladus-Dione pair to get trapped in the 2:1 resonance.

This orbital evolution path is promising to reproduce the current orbits of Enceladus, Tethys, and Dione.
However, in the calculations consisting only of Enceladus,
Tethys and Dione with a constant ring mass, 
we only found the orbital evolution path if Enceladus' eccentricity is
artificially decreased after the Tethys-Dione 2:1 resonance is passed; this is because
Enceladus' eccentricity necessary to bypass the Tethys-Dione 3:2 resonance is so large that 
it results in collisions between Enceladus and Tethys afterward.
Although we assumed that the ring mass is constant to focus ourselves on
orbital evolution, it must evolve with time.
The ring mass decrease due to Mimas formation could 
lower Enceladus' eccentricity to
avoid the collision between Enceladus and Tethys.s
The dynamical effect of Mimas, the smallest moon among the mid-sized moons,
could also play an important role, although we have not
explored the effect. 
The assumption of constant $k_{\rm 2m}/Q_{\rm m}$ for all the moons is
also too simple.
The parameter survey taking these effects into account is left for a separate paper.
 
We also estimated the heat flux of individual moons, which is
caused by thermal energy stored in past periods of high eccentricity
as a result of resonant interactions and ring torque.
Unless Enceladus and Tethys are captured by the mean-motion resonance, 
the heat generation is the highest for Enceladus.
The orbital evolution with the eccentricity excitation
by the ring torque can produce the heat flux
that is comparable to or slightly smaller than the observationally inferred value of Enceladus.
Tidal heating due to the high eccentricity events may
make the moons more convective and dissipative,
which may significantly increase $k_{\rm 2m}/Q_{\rm m}$.
We will also address these issues 
in a separate paper. 

In conclusion, if we take into account of the orbital expansion and eccentricity excitation by ring torque,
there will be one possible pass to solve the problem of the current heat budged on Enceladus and 
the resonance capture from the birth to the current orbit.

\section*{Acknowledgments}
We thank anonymous referees for their helpful and very detailed comments.
This work was supported by JSPS Kakenhi grant 15H02065 and 17K05635.
We thank Daigo Shoji for helpful comments.

\appendix
\section{Equilibrium eccentricity of a resonant pair}

We consider a system of an outwardly migrating planet
in a circular orbit and a test particle 
trapped at the exterior $j:j + 1$ mean-motion resonance with the planet.
According to the outward migration of the planet,
the trapped test particle also migrates outward
and its eccentricity is secularly increased. 
The test particle's eccentricity increase rate is given by \citep{Malhotra1995}
\begin{eqnarray}
\label{eq: e2}
\frac{de^2}{dt} \simeq \frac{1}{j + 1}\frac{1}{a}\frac{da}{dt},
\end{eqnarray}
where $a$ is the semi-major axis of the test particle.
From this relation, it is suggested that for a 
convergent resonant pair of bodies with mass $M_i$,
semi-major axis $a_i$ and eccentricity $e_i$ ($i = 1,2$) 
(the inner body 1 migrates faster than the outer body 2; $ M_1/a_1^{13/2} > M_2/a_2^{13/2}$),
their eccentricity increase rates are 
\begin{eqnarray}
\label{eq: e2sat}
\frac{de_{1}^2}{dt}&\simeq& 
\frac{M_{2}}{M_{1} + M_{2}}\frac{1}{j + 1}\left(\frac{1}{a_{1}}\frac{da_{1}}{dt} - \frac{1}{a_{2}}\frac{da_{2}}{dt} \right), \\
\label{eq: e2sat_2}
\frac{de_{2}^2}{dt} &\simeq&
 \frac{M_{1}}{M_{1} + M_{2}}\frac{1}{j + 1}\left(\frac{1}{a_{1}}\frac{da_{1}}{dt} - \frac{1}{a_{2}}\frac{da_{2}}{dt} \right),
\end{eqnarray}
where
\begin{eqnarray}
\label{eq: where_a}
\frac{1}{a_{1}}\frac{da_{1}}{dt} - \frac{1}{a_{2}}\frac{da_{2}}{dt}= 
3\frac{k_{\rm 2p}}{Q_{\rm p}}\frac{R_{\rm p}^5}{M_{\rm p}}\left(\frac{M_{1}}{a_{1}^5}\Omega_{1} - \frac{M_{2}}{a_{2}^5}\Omega_{2}\right) > 0.
\end{eqnarray}
We neglected the second term in Eq.~(\ref{eq:a_evol2})
for simplicity, 
because it is smaller than the first term for $e \sim 0.03$ 
that we consider here (Eq.~(\ref{eq:a_evol3})).

The eccentricity damping rates by tide are
\begin{eqnarray}
\label{eq: e_damp1}
\frac{de_{1}^2}{dt} &\simeq&
 -21\frac{k_{\rm 2m,1}}{Q_{\rm m,1}}\frac{M_{\rm p}}{M_1}\frac{R_{\rm m,1}}{a_1^5}\Omega_{\rm 1}e_1^2, \\
\label{eq: e_damp2}
\frac{de_{2}^2}{dt} &\simeq& 
-21\frac{k_{\rm 2m,2}}{Q_{\rm m,2}}\frac{M_{\rm p}}{M_2}\frac{R_{\rm m,2}}{a_2^5}\Omega_2e_2^2.
\end{eqnarray}
By balancing (\ref{eq: e2sat}) and (\ref{eq: e2sat_2}) with (\ref{eq: e_damp1}) and (\ref{eq: e_damp2}), $e$ has equilibrium values, 
\begin{eqnarray}
\label{eq: e_equi}
e_1^2 &\sim&
 \frac{M_2}{M_1 + M_2}\frac{1}{7(j + 1)}\frac{k_{\rm 2p}/Q_{\rm p}}{k_{\rm 2m,1}/Q_{\rm m,1}}\left(\frac{M_1}{M_{\rm p}}\right)^2\left(\frac{R_{\rm p}}{R_{\rm m,1}}\right)^5\left[ 1 - \frac{M_2}{M_1}\left(\frac{a_1}{a_2}\right)^{13/2}\right] \nonumber \\
&\sim& 
\frac{M_2}{M_1 + M_2}\frac{1}{7(j + 1)}\frac{k_{\rm 2p}/Q_{\rm p}}{k_{\rm 2m,1}/Q_{\rm m,1}}\frac{R_{\rm m,1}}{R_{\rm p}}\left[ 1 - \frac{M_2}{M_1}\left(\frac{j}{j + 1}\right)^{13/3}\right], \\
e_2^2 &\sim& 
\frac{M_1}{M_1 + M_2}\frac{1}{7(j + 1)}\frac{k_{\rm 2p}/Q_{\rm p}}{k_{\rm 2m,2}/Q_{\rm m,2}}\frac{R_{\rm m,2}}{R_{\rm p}}\left[ \frac{M_1}{M_2}\left(\frac{j+1}{j }\right)^{13/3} - 1\right].
\end{eqnarray}
For the Enceladus and Dione pair trapped at 2:1 resonance,
$j = 1$ and body 1 and 2 are Enceladus and Dione, respectively.
Substituting $ M_1/M_2 = M_{\rm E}/M_{\rm D} \sim 0.1$, $R_{\rm E}/R_{\rm p} \sim 0.0042$ 
and $R_{\rm D}/R_{\rm p} \sim 0.0093$ into the above equations, we obtain
\begin{eqnarray}
\label{eq: e_enc}
e_{\rm{E}} &\sim& 
0.038 \left( \frac{k_{\rm 2p}/Q_{\rm p}}{10^{-4}} \right)^{1/2} \left( \frac{k_{\rm 2m}/Q_{\rm m}}{10^{-5}} \right)^{-1/2}, \\
e_{\rm{D}} &\sim& 
0.025 \left( \frac{k_{\rm 2p}/Q_{\rm p}}{10^{-4}} \right)^{1/2} \left( \frac{k_{\rm 2m}/Q_{\rm m}}{10^{-5}} \right)^{-1/2}, 
\end{eqnarray}
where we used $k_{\rm 2p} \sim 0.3$, $Q_{\rm p} \sim 3000$, 
$k_{\rm 2m} \sim 10^{-3}$ and $Q_{\rm m} \sim 10^2$. 
These eccentricities are one order higher than the current values, 
one suggestion is raised that their eccentricities are now on the way to the equilibrium.

Note that the derivation for the equilibrium eccentricity here 
is simplified. 
More rigorous derivations with Lagrange equations and detailed resonant properties
are found in the past literatures \citep[e.g.,][]{Meyer2008a,Zhang2009}.
While the numerical factors differ from the rigorous treatment
by a factor of up to a few, the dependence on $k_{\rm 2,p}/Q_{\rm p}$ and
$k_{\rm 2,m}/Q_{\rm m}$ are reproduced by the simple derivation here
and the difference in the numerical factors does not affect 
the discussions in this paper.

\section*{References}

\bibliography{mybibfile}

\begin{thebibliography}{39}
\expandafter\ifx\csname natexlab\endcsname\relax\def\natexlab#1{#1}\fi
\providecommand{\url}[1]{\texttt{#1}}
\providecommand{\href}[2]{#2}
\providecommand{\path}[1]{#1}
\providecommand{\DOIprefix}{doi:}
\providecommand{\ArXivprefix}{arXiv:}
\providecommand{\URLprefix}{URL: }
\providecommand{\Pubmedprefix}{pmid:}
\providecommand{\doi}[1]{\href{http://dx.doi.org/#1}{\path{#1}}}
\providecommand{\Pubmed}[1]{\href{pmid:#1}{\path{#1}}}
\providecommand{\bibinfo}[2]{#2}
\ifx\xfnm\relax \def\xfnm[#1]{\unskip,\space#1}\fi
\bibitem[{Asphaug et~al.(2006)Asphaug, Agnor \& Williams}]{Asphaug2006}
\bibinfo{author}{Asphaug, E.}, \bibinfo{author}{Agnor, C.~B.}, \&
  \bibinfo{author}{Williams, Q.} (\bibinfo{year}{2006}).
\newblock \bibinfo{title}{{Hit-and-run planetary collisions}}.
\newblock {\it \bibinfo{journal}{Nature}\/},  {\it \bibinfo{volume}{439}\/},
  \bibinfo{pages}{155--160}. \DOIprefix\doi{10.1038/nature04311}.
\bibitem[{Charnoz et~al.(2011)Charnoz, Crida, Castillo-Rogez, Lainey, Dones,
  Karatekin, Tobie, Mathis, {Le Poncin-Lafitte} \& Salmon}]{Charnoz2011}
\bibinfo{author}{Charnoz, S.}, \bibinfo{author}{Crida, A.},
  \bibinfo{author}{Castillo-Rogez, J.~C.}, \bibinfo{author}{Lainey, V.},
  \bibinfo{author}{Dones, L.}, \bibinfo{author}{Karatekin, {\"{O}}.},
  \bibinfo{author}{Tobie, G.}, \bibinfo{author}{Mathis, S.},
  \bibinfo{author}{{Le Poncin-Lafitte}, C.}, \& \bibinfo{author}{Salmon, J.}
  (\bibinfo{year}{2011}).
\newblock \bibinfo{title}{{Accretion of Saturn's mid-sized moons during the
  viscous spreading of young massive rings: Solving the paradox of
  silicate-poor rings versus silicate-rich moons}}.
\newblock {\it \bibinfo{journal}{Icarus}\/},  {\it \bibinfo{volume}{216}\/},
  \bibinfo{pages}{535--550}. \DOIprefix\doi{10.1016/j.icarus.2011.09.017}.
  \href{http://arxiv.org/abs/1109.3360}{\tt arXiv:1109.3360}.
\bibitem[{{Chen} \& {Nimmo}(2008)}]{Chen2008}
\bibinfo{author}{{Chen}, E.~M.~A.}, \& \bibinfo{author}{{Nimmo}, F.}
  (\bibinfo{year}{2008}).
\newblock \bibinfo{title}{{Implications from Ithaca Chasma for the thermal and
  orbital history of Tethys}}.
\newblock {\it \bibinfo{journal}{grl}\/},  {\it \bibinfo{volume}{35}\/},
  \bibinfo{pages}{L19203}. \DOIprefix\doi{10.1029/2008GL035402}.
\bibitem[{{Choblet} et~al.(2017){Choblet}, {Tobie}, {Sotin}, {B{\v
  e}hounkov{\'a}}, {{\v C}adek}, {Postberg} \& {Sou{\v c}ek}}]{Choblet2017}
\bibinfo{author}{{Choblet}, G.}, \bibinfo{author}{{Tobie}, G.},
  \bibinfo{author}{{Sotin}, C.}, \bibinfo{author}{{B{\v e}hounkov{\'a}}, M.},
  \bibinfo{author}{{{\v C}adek}, O.}, \bibinfo{author}{{Postberg}, F.}, \&
  \bibinfo{author}{{Sou{\v c}ek}, O.} (\bibinfo{year}{2017}).
\newblock \bibinfo{title}{{Powering prolonged hydrothermal activity inside
  Enceladus}}.
\newblock {\it \bibinfo{journal}{Nature Astronomy}\/},  {\it
  \bibinfo{volume}{1}\/}, \bibinfo{pages}{841--847}.
  \DOIprefix\doi{10.1038/s41550-017-0289-8}.
\bibitem[{Crida \& Charnoz(2012)}]{Crida2012}
\bibinfo{author}{Crida, a.}, \& \bibinfo{author}{Charnoz, S.}
  (\bibinfo{year}{2012}).
\newblock \bibinfo{title}{{Formation of regular satellites from ancient massive
  rings in the solar system.}}
\newblock {\it \bibinfo{journal}{Science (New York, N.Y.)}\/},  {\it
  \bibinfo{volume}{338}\/}, \bibinfo{pages}{1196--9}. \URLprefix
  \url{http://www.ncbi.nlm.nih.gov/pubmed/23197530}.
  \DOIprefix\doi{10.1126/science.1226477}.
  \href{http://arxiv.org/abs/1301.3808}{\tt arXiv:1301.3808}.
\bibitem[{Duffell \& Chiang(2015)}]{Duffell2015}
\bibinfo{author}{Duffell, P.~C.}, \& \bibinfo{author}{Chiang, E.}
  (\bibinfo{year}{2015}).
\newblock \bibinfo{title}{{Eccentric Jupiters Via Disk-Planet Interactions}}.
\newblock {\it \bibinfo{journal}{Astrophysical Journal}\/},  {\it
  \bibinfo{volume}{812}\/}, \bibinfo{pages}{1DUMMY}. \URLprefix
  \url{http://dx.doi.org/10.1088/0004-637X/812/2/94}.
  \DOIprefix\doi{10.1088/0004-637X/812/2/94}.
  \href{http://arxiv.org/abs/1507.08667}{\tt arXiv:1507.08667}.
\bibitem[{Duncan et~al.(1998)Duncan, Levison \& Lee}]{Duncan1998}
\bibinfo{author}{Duncan, M.~J.}, \bibinfo{author}{Levison, H.~F.}, \&
  \bibinfo{author}{Lee, M.~H.} (\bibinfo{year}{1998}).
\newblock \bibinfo{title}{{A Multiple Time Step Symplectic Algorithm for
  Integrating Close Encounters}}.
\newblock {\it \bibinfo{journal}{The Astronomical Journal}\/},  {\it
  \bibinfo{volume}{116}\/}, \bibinfo{pages}{2067--2077}. \URLprefix
  \url{http://adsabs.harvard.edu/cgi-bin/nph-data{\_}query?bibcode=1998AJ....116.2067D{\&}link{\_}type=ABSTRACT{\%}5Cnpapers3://publication/doi/10.1086/300541}.
  \DOIprefix\doi{10.1086/300541}.
\bibitem[{{Ferraz-Mello} et~al.(2017){Ferraz-Mello}, {Folonier} \&
  {Andrade-Ines}}]{Ferraz-Mello2017}
\bibinfo{author}{{Ferraz-Mello}, S.}, \bibinfo{author}{{Folonier}, H.~A.}, \&
  \bibinfo{author}{{Andrade-Ines}, E.} (\bibinfo{year}{2017}).
\newblock \bibinfo{title}{{Tidal synchronization of close-in satellites and
  exoplanets. III. Tidal dissipation revisited and application to Enceladus}}.
\newblock {\it \bibinfo{journal}{ArXiv e-prints}\/}, .
  \href{http://arxiv.org/abs/1707.09229}{\tt arXiv:1707.09229}.
\bibitem[{Fuller et~al.(2016)Fuller, Luan \& Quataert}]{Fuller2016}
\bibinfo{author}{Fuller, J.}, \bibinfo{author}{Luan, J.}, \&
  \bibinfo{author}{Quataert, E.} (\bibinfo{year}{2016}).
\newblock \bibinfo{title}{{Resonance locking as the source of rapid tidal
  migration in the Jupiter and Saturn moon systems}}.
\newblock {\it \bibinfo{journal}{Monthly Notices of the Royal Astronomical
  Society}\/},  {\it \bibinfo{volume}{458}\/}, \bibinfo{pages}{3867--3879}.
  \DOIprefix\doi{10.1093/mnras/stw609}.
  \href{http://arxiv.org/abs/1601.05804}{\tt arXiv:1601.05804}.
\bibitem[{Gavrilov \& Zharkov(1977)}]{Gavrilov1977}
\bibinfo{author}{Gavrilov, S.~V.}, \& \bibinfo{author}{Zharkov, V.~N.}
  (\bibinfo{year}{1977}).
\newblock \bibinfo{title}{{Love numbers of the giant planets}}.
\newblock {\it \bibinfo{journal}{Icarus}\/},  {\it \bibinfo{volume}{32}\/},
  \bibinfo{pages}{443--449}. \DOIprefix\doi{10.1016/0019-1035(77)90015-X}.
\bibitem[{Genda et~al.(2012)Genda, Kokubo \& Ida}]{Genda2012}
\bibinfo{author}{Genda, H.}, \bibinfo{author}{Kokubo, E.}, \&
  \bibinfo{author}{Ida, S.} (\bibinfo{year}{2012}).
\newblock \bibinfo{title}{{Merging Criteria for Giant Impacts of
  Protoplanets}}.
\newblock {\it \bibinfo{journal}{The Astrophysical Journal}\/},  {\it
  \bibinfo{volume}{744}\/}, \bibinfo{pages}{137}. \URLprefix
  \url{http://stacks.iop.org/0004-637X/744/i=2/a=137?key=crossref.13660ad625848d1eee511669b9db133a}.
  \DOIprefix\doi{10.1088/0004-637X/744/2/137}.
  \href{http://arxiv.org/abs/arXiv:1109.4330v1}{\tt arXiv:arXiv:1109.4330v1}.
\bibitem[{{Giese} et~al.(2007){Giese}, {Wagner}, {Neukum}, {Helfenstein} \&
  {Thomas}}]{Giese2007}
\bibinfo{author}{{Giese}, B.}, \bibinfo{author}{{Wagner}, R.},
  \bibinfo{author}{{Neukum}, G.}, \bibinfo{author}{{Helfenstein}, P.}, \&
  \bibinfo{author}{{Thomas}, P.~C.} (\bibinfo{year}{2007}).
\newblock \bibinfo{title}{{Tethys: Lithospheric thickness and heat flux from
  flexurally supported topography at Ithaca Chasma}}.
\newblock {\it \bibinfo{journal}{grl}\/},  {\it \bibinfo{volume}{34}\/},
  \bibinfo{pages}{L21203}. \DOIprefix\doi{10.1029/2007GL031467}.
\bibitem[{Goldreich \& Sari(2003)}]{Goldreich2003b}
\bibinfo{author}{Goldreich, P.}, \& \bibinfo{author}{Sari, R.}
  (\bibinfo{year}{2003}).
\newblock \bibinfo{title}{{Eccentricity Evolution for Planets in Gaseous
  Disks}}.
\newblock {\it \bibinfo{journal}{The Astrophysical Journal}\/},  {\it
  \bibinfo{volume}{585}\/}, \bibinfo{pages}{1024--1037}. \URLprefix
  \url{http://stacks.iop.org/0004-637X/585/i=2/a=1024}.
  \DOIprefix\doi{10.1086/346202}. \href{http://arxiv.org/abs/0202462}{\tt
  arXiv:0202462}.
\bibitem[{Goldreich \& Soter(1966)}]{Goldreich1966}
\bibinfo{author}{Goldreich, P.}, \& \bibinfo{author}{Soter, S.}
  (\bibinfo{year}{1966}).
\newblock \bibinfo{title}{{Q in the solar system}}.
\newblock {\it \bibinfo{journal}{Icarus}\/},  {\it \bibinfo{volume}{5}\/},
  \bibinfo{pages}{375--389}. \URLprefix
  \url{http://linkinghub.elsevier.com/retrieve/pii/0019103566900510}.
  \DOIprefix\doi{10.1016/0019-1035(66)90051-0}.
\bibitem[{Helled \& Guillot(2013)}]{Helled2013}
\bibinfo{author}{Helled, R.}, \& \bibinfo{author}{Guillot, T.}
  (\bibinfo{year}{2013}).
\newblock \bibinfo{title}{{Interior models of saturn: Including the
  uncertainties in shape and rotation}}.
\newblock {\it \bibinfo{journal}{Astrophysical Journal}\/},  {\it
  \bibinfo{volume}{767}\/}. \DOIprefix\doi{10.1088/0004-637X/767/2/113}.
  \href{http://arxiv.org/abs/arXiv:1302.6690v1}{\tt arXiv:arXiv:1302.6690v1}.
\bibitem[{Howett et~al.(2011)Howett, Spencer, Pearl \& Segura}]{Howett2011}
\bibinfo{author}{Howett, C. J.~A.}, \bibinfo{author}{Spencer, J.~R.},
  \bibinfo{author}{Pearl, J.}, \& \bibinfo{author}{Segura, M.}
  (\bibinfo{year}{2011}).
\newblock \bibinfo{title}{{High heat flow from Enceladus' south polar region
  measured using 10-600 cm-1 Cassini/CIRS data}}.
\newblock {\it \bibinfo{journal}{Journal of Geophysical Research E:
  Planets}\/},  {\it \bibinfo{volume}{116}\/}, \bibinfo{pages}{1--15}.
  \DOIprefix\doi{10.1029/2010JE003718}.
\bibitem[{{Ida}(1990)}]{Ida1990}
\bibinfo{author}{{Ida}, S.} (\bibinfo{year}{1990}).
\newblock \bibinfo{title}{{Stirring and dynamical friction rates of
  planetesimals in the solar gravitational field}}.
\newblock {\it \bibinfo{journal}{Icarus}\/},  {\it \bibinfo{volume}{88}\/},
  \bibinfo{pages}{129--145}. \DOIprefix\doi{10.1016/0019-1035(90)90182-9}.
\bibitem[{Ida \& Nakazawa(1989)}]{Ida1989}
\bibinfo{author}{Ida, S.}, \& \bibinfo{author}{Nakazawa, K.}
  (\bibinfo{year}{1989}).
\newblock \bibinfo{title}{{Collisional probability of planetesimals revolving
  in the solar gravitational field. III}}.
\newblock {\it \bibinfo{journal}{Astronomy and Astrophysics (ISSN
  0004-6361)}\/},  {\it \bibinfo{volume}{224}\/}, \bibinfo{pages}{303--315}.
  \URLprefix
  \url{http://adsabs.harvard.edu/abs/1989A{\&}A...224..303I{\%}5Cnpapers3://publication/uuid/6C8ECD9B-6E4A-4F17-8D88-CF14E571F7C3}.
  \DOIprefix\doi{10.1017/CBO9781107415324.004}.
  \href{http://arxiv.org/abs/arXiv:1011.1669v3}{\tt arXiv:arXiv:1011.1669v3}.
\bibitem[{Kominami \& Ida(2002)}]{Kominami2002}
\bibinfo{author}{Kominami, J.}, \& \bibinfo{author}{Ida, S.}
  (\bibinfo{year}{2002}).
\newblock \bibinfo{title}{{The Effect of Tidal Interaction with a Gas Disk on
  Formation of Terrestrial Planets}}.
\newblock {\it \bibinfo{journal}{Icarus}\/},  {\it \bibinfo{volume}{157}\/},
  \bibinfo{pages}{43--56}. \URLprefix
  \url{http://linkinghub.elsevier.com/retrieve/doi/10.1006/icar.2001.6811}.
  \DOIprefix\doi{10.1006/icar.2001.6811}.
\bibitem[{Lainey et~al.(2017)Lainey, Jacobson, Tajeddine, Cooper, Murray,
  Robert, Tobie, Guillot, Mathis, Remus, Desmars, Arlot, {De Cuyper}, Dehant,
  Pascu, Thuillot, Poncin-Lafitte \& Zahn}]{Lainey2017}
\bibinfo{author}{Lainey, V.}, \bibinfo{author}{Jacobson, R.~A.},
  \bibinfo{author}{Tajeddine, R.}, \bibinfo{author}{Cooper, N.~J.},
  \bibinfo{author}{Murray, C.}, \bibinfo{author}{Robert, V.},
  \bibinfo{author}{Tobie, G.}, \bibinfo{author}{Guillot, T.},
  \bibinfo{author}{Mathis, S.}, \bibinfo{author}{Remus, F.},
  \bibinfo{author}{Desmars, J.}, \bibinfo{author}{Arlot, J.~E.},
  \bibinfo{author}{{De Cuyper}, J.~P.}, \bibinfo{author}{Dehant, V.},
  \bibinfo{author}{Pascu, D.}, \bibinfo{author}{Thuillot, W.},
  \bibinfo{author}{Poncin-Lafitte, C.~L.}, \& \bibinfo{author}{Zahn, J.~P.}
  (\bibinfo{year}{2017}).
\newblock \bibinfo{title}{{New constraints on Saturn's interior from Cassini
  astrometric data}}.
\newblock {\it \bibinfo{journal}{Icarus}\/},  {\it \bibinfo{volume}{281}\/},
  \bibinfo{pages}{286--296}. \DOIprefix\doi{10.1016/j.icarus.2016.07.014}.
  \href{http://arxiv.org/abs/1510.05870}{\tt arXiv:1510.05870}.
\bibitem[{Lainey et~al.(2012)Lainey, Karatekin, Desmars, Charnoz, Arlot,
  Emelyanov, {Le Poncin-Lafitte}, Mathis, Remus, Tobie \& Zahn}]{Lainey2012}
\bibinfo{author}{Lainey, V.}, \bibinfo{author}{Karatekin, {\"{O}}.},
  \bibinfo{author}{Desmars, J.}, \bibinfo{author}{Charnoz, S.},
  \bibinfo{author}{Arlot, J.-E.}, \bibinfo{author}{Emelyanov, N.},
  \bibinfo{author}{{Le Poncin-Lafitte}, C.}, \bibinfo{author}{Mathis, S.},
  \bibinfo{author}{Remus, F.}, \bibinfo{author}{Tobie, G.}, \&
  \bibinfo{author}{Zahn, J.-P.} (\bibinfo{year}{2012}).
\newblock \bibinfo{title}{{Strong Tidal Dissipation in Saturn and Constraints
  on Enceladus' Thermal State From Astrometry}}.
\newblock {\it \bibinfo{journal}{The Astrophysical Journal}\/},  {\it
  \bibinfo{volume}{752}\/}, \bibinfo{pages}{14}. \URLprefix
  \url{http://stacks.iop.org/0004-637X/752/i=1/a=14?key=crossref.d68a903d62213eb4f281cc2f4669349f}.
  \DOIprefix\doi{10.1088/0004-637X/752/1/14}.
  \href{http://arxiv.org/abs/1204.0895}{\tt arXiv:1204.0895}.
\bibitem[{Malhotra(1993)}]{Malhotra1993}
\bibinfo{author}{Malhotra, R.} (\bibinfo{year}{1993}).
\newblock \bibinfo{title}{{Orbital resonances in the solar nebula - Strengths
  and weaknesses}}.
\newblock {\it \bibinfo{journal}{Icarus}\/},  {\it \bibinfo{volume}{106}\/},
  \bibinfo{pages}{264}. \DOIprefix\doi{10.1006/icar.1993.1170}.
\bibitem[{Malhotra(1995)}]{Malhotra1995}
\bibinfo{author}{Malhotra, R.} (\bibinfo{year}{1995}).
\newblock \bibinfo{title}{{The Origin of Pluto's Orbit: Implications for the
  Solar System Beyond Neptune}}.
\newblock {\it \bibinfo{journal}{Astronomical Journal}\/},  {\it
  \bibinfo{volume}{110}\/}, \bibinfo{pages}{420}.
  \DOIprefix\doi{10.1086/117532}. \href{http://arxiv.org/abs/9504036}{\tt
  arXiv:9504036}.
\bibitem[{Malhotra \& Dermott(1990)}]{Malhotra1990}
\bibinfo{author}{Malhotra, R.}, \& \bibinfo{author}{Dermott, S.~F.}
  (\bibinfo{year}{1990}).
\newblock \bibinfo{title}{{The role of secondary resonances in the orbital
  history of Miranda}}.
\newblock {\it \bibinfo{journal}{Icarus}\/},  {\it \bibinfo{volume}{85}\/},
  \bibinfo{pages}{444--480}. \DOIprefix\doi{10.1016/0019-1035(90)90126-T}.
\bibitem[{Meyer \& Wisdom(2007)}]{Meyer2007}
\bibinfo{author}{Meyer, J.}, \& \bibinfo{author}{Wisdom, J.}
  (\bibinfo{year}{2007}).
\newblock \bibinfo{title}{{Tidal heating in Enceladus}}.
\newblock {\it \bibinfo{journal}{Icarus}\/},  {\it \bibinfo{volume}{188}\/},
  \bibinfo{pages}{535--539}. \DOIprefix\doi{10.1016/j.icarus.2007.03.001}.
\bibitem[{{Meyer} \& {Wisdom}(2008)}]{Meyer2008b}
\bibinfo{author}{{Meyer}, J.}, \& \bibinfo{author}{{Wisdom}, J.}
  (\bibinfo{year}{2008}).
\newblock \bibinfo{title}{{Episodic volcanism on Enceladus: Application of the
  Ojakangas Stevenson model}}.
\newblock {\it \bibinfo{journal}{Icarus}\/},  {\it \bibinfo{volume}{198}\/},
  \bibinfo{pages}{178--180}. \DOIprefix\doi{10.1016/j.icarus.2008.06.012}.
\bibitem[{Meyer \& Wisdom(2008)}]{Meyer2008a}
\bibinfo{author}{Meyer, J.}, \& \bibinfo{author}{Wisdom, J.}
  (\bibinfo{year}{2008}).
\newblock \bibinfo{title}{{Tidal evolution of Mimas, Enceladus, and Dione}}.
\newblock {\it \bibinfo{journal}{Icarus}\/},  {\it \bibinfo{volume}{193}\/},
  \bibinfo{pages}{213--223}. \DOIprefix\doi{10.1016/j.icarus.2007.09.008}.
\bibitem[{{Meyer-Vernet} \& {Sicardy}(1987)}]{Meyer-Vernet1987}
\bibinfo{author}{{Meyer-Vernet}, N.}, \& \bibinfo{author}{{Sicardy}, B.}
  (\bibinfo{year}{1987}).
\newblock \bibinfo{title}{{On the physics of resonant disk-satellite
  interaction}}.
\newblock {\it \bibinfo{journal}{Icarus}\/},  {\it \bibinfo{volume}{69}\/},
  \bibinfo{pages}{157--175}. \DOIprefix\doi{10.1016/0019-1035(87)90011-X}.
\bibitem[{{Murray} \& {Dermott}(1999)}]{MurrayDermott1999}
\bibinfo{author}{{Murray}, C.~D.}, \& \bibinfo{author}{{Dermott}, S.~F.}
  (\bibinfo{year}{1999}).
\newblock {\it \bibinfo{title}{{Solar system dynamics}}\/}.
\bibitem[{Ojakangas \& Stevenson(1986)}]{Ojakangas1986}
\bibinfo{author}{Ojakangas, G.~W.}, \& \bibinfo{author}{Stevenson, D.~J.}
  (\bibinfo{year}{1986}).
\newblock \bibinfo{title}{{Episodic volcanism of tidally heated satellites with
  application to Io}}.
\newblock {\it \bibinfo{journal}{Icarus}\/},  {\it \bibinfo{volume}{66}\/},
  \bibinfo{pages}{341--358}. \DOIprefix\doi{10.1016/0019-1035(86)90163-6}.
\bibitem[{O'Neill \& Nimmo(2010)}]{ONeill2010}
\bibinfo{author}{O'Neill, C.}, \& \bibinfo{author}{Nimmo, F.}
  (\bibinfo{year}{2010}).
\newblock \bibinfo{title}{{The role of episodic overturn in generating the
  surface geology and heat flow on Enceladus}}.
\newblock {\it \bibinfo{journal}{Nature Geoscience}\/},  {\it
  \bibinfo{volume}{3}\/}, \bibinfo{pages}{88--91}. \URLprefix
  \url{http://dx.doi.org/10.1038/ngeo731}. \DOIprefix\doi{10.1038/ngeo731}.
\bibitem[{Peale et~al.(1980)Peale, Cassen \& Reynolds}]{Peale1980a}
\bibinfo{author}{Peale, S.~J.}, \bibinfo{author}{Cassen, P.}, \&
  \bibinfo{author}{Reynolds, R.} (\bibinfo{year}{1980}).
\newblock \bibinfo{title}{{Tidal dissipation, orbital evolution, and the nature
  of Saturn's inner satellites}}.
\newblock {\it \bibinfo{journal}{Icarus}\/},  {\it \bibinfo{volume}{72}\/},
  \bibinfo{pages}{1196--1205}. \URLprefix
  \url{http://www.sciencedirect.com/science/article/pii/0019103580900883}.
  \DOIprefix\doi{10.1016/0019-1035(80)90088-3}.
\bibitem[{{Petit} \& {Henon}(1986)}]{Petit1986}
\bibinfo{author}{{Petit}, J.-M.}, \& \bibinfo{author}{{Henon}, M.}
  (\bibinfo{year}{1986}).
\newblock \bibinfo{title}{{Satellite encounters}}.
\newblock {\it \bibinfo{journal}{Icarus}\/},  {\it \bibinfo{volume}{66}\/},
  \bibinfo{pages}{536--555}. \DOIprefix\doi{10.1016/0019-1035(86)90089-8}.
\bibitem[{Porco et~al.(2006)Porco, Helfenstein, Thomas, Ingersoll, Wisdom,
  West, Neukum, Denk, Wagner, Roatsch, Kieffer, Turtle, McEwen, Johnson,
  Rathbun, Veverka, Wilson, Perry, Spitale, Brahic, Burns, DelGenio, Dones,
  Murray \& Squyres}]{Porco2006}
\bibinfo{author}{Porco, C.~C.}, \bibinfo{author}{Helfenstein, P.},
  \bibinfo{author}{Thomas, P.~C.}, \bibinfo{author}{Ingersoll, A.~P.},
  \bibinfo{author}{Wisdom, J.}, \bibinfo{author}{West, R.},
  \bibinfo{author}{Neukum, G.}, \bibinfo{author}{Denk, T.},
  \bibinfo{author}{Wagner, R.}, \bibinfo{author}{Roatsch, T.},
  \bibinfo{author}{Kieffer, S.}, \bibinfo{author}{Turtle, E.},
  \bibinfo{author}{McEwen, A.}, \bibinfo{author}{Johnson, T.~V.},
  \bibinfo{author}{Rathbun, J.}, \bibinfo{author}{Veverka, J.},
  \bibinfo{author}{Wilson, D.}, \bibinfo{author}{Perry, J.},
  \bibinfo{author}{Spitale, J.}, \bibinfo{author}{Brahic, A.},
  \bibinfo{author}{Burns, J.~A.}, \bibinfo{author}{DelGenio, A.~D.},
  \bibinfo{author}{Dones, L.}, \bibinfo{author}{Murray, C.~D.}, \&
  \bibinfo{author}{Squyres, S.} (\bibinfo{year}{2006}).
\newblock \bibinfo{title}{{Cassini observes the active south pole of
  enceladus}}.
\newblock {\it \bibinfo{journal}{Science}\/},  {\it \bibinfo{volume}{311}\/},
  \bibinfo{pages}{1393--1401}. \DOIprefix\doi{10.1126/science.1123013}.
\bibitem[{Salmon \& Canup(2017)}]{Salmon2017}
\bibinfo{author}{Salmon, J.}, \& \bibinfo{author}{Canup, R.~M.}
  (\bibinfo{year}{2017}).
\newblock \bibinfo{title}{{Accretion of Saturn's Inner Mid-sized Moons from a
  Massive Primordial Ice Ring}}, .
\newblock (pp. \bibinfo{pages}{1--28}). \URLprefix
  \url{http://arxiv.org/abs/1702.04385{\%}0Ahttp://dx.doi.org/10.3847/1538-4357/836/1/109}.
  \DOIprefix\doi{10.3847/1538-4357/836/1/109}.
  \href{http://arxiv.org/abs/1702.04385}{\tt arXiv:1702.04385}.
\bibitem[{Showman et~al.(1997)Showman, Stevenson \& Malhotra}]{showman1997}
\bibinfo{author}{Showman, A.~P.}, \bibinfo{author}{Stevenson, D.~J.}, \&
  \bibinfo{author}{Malhotra, R.} (\bibinfo{year}{1997}).
\newblock \bibinfo{title}{{Coupled Orbital and Thermal Evolution of Ganymede}}.
\newblock {\it \bibinfo{journal}{Icarus}\/},  {\it \bibinfo{volume}{129}\/},
  \bibinfo{pages}{367--383}. \URLprefix
  \url{http://linkinghub.elsevier.com/retrieve/pii/S001910359795778X}.
  \DOIprefix\doi{10.1006/icar.1997.5778}.
\bibitem[{Spencer \& Nimmo(2013)}]{Spencer2013}
\bibinfo{author}{Spencer, J.~R.}, \& \bibinfo{author}{Nimmo, F.}
  (\bibinfo{year}{2013}).
\newblock \bibinfo{title}{{Enceladus: An Active Ice World in the Saturn
  System}}.
\newblock {\it \bibinfo{journal}{Annual Review of Earth and Planetary
  Sciences}\/},  {\it \bibinfo{volume}{41}\/}, \bibinfo{pages}{693--717}.
  \URLprefix
  \url{http://www.annualreviews.org/doi/10.1146/annurev-earth-050212-124025}.
  \DOIprefix\doi{10.1146/annurev-earth-050212-124025}.
\bibitem[{Zhang \& Nimmo(2009)}]{Zhang2009}
\bibinfo{author}{Zhang, K.}, \& \bibinfo{author}{Nimmo, F.}
  (\bibinfo{year}{2009}).
\newblock \bibinfo{title}{{Recent orbital evolution and the internal structures
  of Enceladus and Dione}}.
\newblock {\it \bibinfo{journal}{Icarus}\/},  {\it \bibinfo{volume}{204}\/},
  \bibinfo{pages}{597--609}. \URLprefix
  \url{http://dx.doi.org/10.1016/j.icarus.2009.07.007}.
  \DOIprefix\doi{10.1016/j.icarus.2009.07.007}.
\bibitem[{Zhang \& Nimmo(2012)}]{Zhang2012}
\bibinfo{author}{Zhang, K.}, \& \bibinfo{author}{Nimmo, F.}
  (\bibinfo{year}{2012}).
\newblock \bibinfo{title}{{Late-stage impacts and the orbital and thermal
  evolution of Tethys}}.
\newblock {\it \bibinfo{journal}{Icarus}\/},  {\it \bibinfo{volume}{218}\/},
  \bibinfo{pages}{348--355}. \URLprefix
  \url{http://dx.doi.org/10.1016/j.icarus.2011.12.013}.
  \DOIprefix\doi{10.1016/j.icarus.2011.12.013}.

\end{thebibliography}

\end{document}